\documentclass[twocolumn,iop,numberedappendix,appendixfloats]{openjournal}
\usepackage[utf8]{inputenc}
\usepackage{amsmath,amssymb,amstext}
\usepackage{graphicx}
\graphicspath{ {./Figures/} }
\usepackage{bm}
\usepackage{natbib}
\usepackage{subcaption}
\usepackage[colorlinks,allcolors=blue]{hyperref}
\usepackage{dsfont}
\usepackage{comment}
\usepackage{float}
\usepackage{xcolor}
\usepackage{listings}
\usepackage{fontawesome}
\definecolor{maroon}{cmyk}{0, 0.87, 0.68, 0.32}
\definecolor{halfgray}{gray}{0.55}
\definecolor{ipython_frame}{RGB}{207, 207, 207}
\definecolor{ipython_bg}{RGB}{247, 247, 247}
\definecolor{ipython_red}{RGB}{186, 33, 33}
\definecolor{ipython_green}{RGB}{0, 128, 0}
\definecolor{ipython_cyan}{RGB}{64, 128, 128}
\definecolor{ipython_purple}{RGB}{170, 34, 255}

\lstdefinelanguage{iPython}{
    morekeywords={access,and,break,class,continue,def,del,elif,else,except,exec,finally,for,from,global,if,import,in,is,lambda,not,or,pass,print,raise,return,try,while},%
    morekeywords=[2]{abs,all,any,basestring,bin,bool,bytearray,callable,chr,classmethod,cmp,compile,complex,delattr,dict,dir,divmod,enumerate,eval,execfile,file,filter,float,format,frozenset,getattr,globals,hasattr,hash,help,hex,id,input,int,isinstance,issubclass,iter,len,list,locals,long,map,max,memoryview,min,next,object,oct,open,ord,pow,property,range,raw_input,reduce,reload,repr,reversed,round,set,setattr,slice,sorted,staticmethod,str,sum,super,tuple,type,unichr,unicode,vars,xrange,zip,apply,buffer,coerce,intern, function, @model, Uniform, Normal, MvNormal, theory_planck},%
    sensitive=true,%
    morecomment=[l]\#,%
    morestring=[b]',%
    morestring=[b]",%
    morestring=[s]{'''}{'''},% used for documentation text (mulitiline strings)
    morestring=[s]{"""}{"""},% added by Philipp Matthias Hahn
    morestring=[s]{r'}{'},% `raw' strings
    morestring=[s]{r"}{"},%
    morestring=[s]{r'''}{'''},%
    morestring=[s]{r"""}{"""},%
    morestring=[s]{u'}{'},% unicode strings
    morestring=[s]{u"}{"},%
    morestring=[s]{u'''}{'''},%
    morestring=[s]{u"""}{"""},%
    literate=
    {á}{{\'a}}1 {é}{{\'e}}1 {í}{{\'i}}1 {ó}{{\'o}}1 {ú}{{\'u}}1
    {Á}{{\'A}}1 {É}{{\'E}}1 {Í}{{\'I}}1 {Ó}{{\'O}}1 {Ú}{{\'U}}1
    {à}{{\`a}}1 {è}{{\`e}}1 {ì}{{\`i}}1 {ò}{{\`o}}1 {ù}{{\`u}}1
    {À}{{\`A}}1 {È}{{\'E}}1 {Ì}{{\`I}}1 {Ò}{{\`O}}1 {Ù}{{\`U}}1
    {ä}{{\"a}}1 {ë}{{\"e}}1 {ï}{{\"i}}1 {ö}{{\"o}}1 {ü}{{\"u}}1
    {Ä}{{\"A}}1 {Ë}{{\"E}}1 {Ï}{{\"I}}1 {Ö}{{\"O}}1 {Ü}{{\"U}}1
    {â}{{\^a}}1 {ê}{{\^e}}1 {î}{{\^i}}1 {ô}{{\^o}}1 {û}{{\^u}}1
    {Â}{{\^A}}1 {Ê}{{\^E}}1 {Î}{{\^I}}1 {Ô}{{\^O}}1 {Û}{{\^U}}1
    {œ}{{\oe}}1 {Œ}{{\OE}}1 {æ}{{\ae}}1 {Æ}{{\AE}}1 {ß}{{\ss}}1
    {ç}{{\c c}}1 {Ç}{{\c C}}1 {ø}{{\o}}1 {å}{{\r a}}1 {Å}{{\r A}}1
    {€}{{\EUR}}1 {£}{{\pounds}}1
    {^}{{{\color{ipython_purple}\^{}}}}1
    {=}{{{\color{ipython_purple}=}}}1
    {+}{{{\color{ipython_purple}+}}}1
    {-}{{{\color{ipython_purple}-}}}1
    {*}{{{\color{ipython_purple}$^\ast$}}}1
    {/}{{{\color{ipython_purple}/}}}1
    {+=}{{{+=}}}1
    {-=}{{{-=}}}1
    {*=}{{{$^\ast$=}}}1
    {/=}{{{/=}}}1,
    literate=
    *{-}{{{\color{ipython_purple}-}}}1
     {?}{{{\color{ipython_purple}?}}}1,
    identifierstyle=\color{black}\ttfamily,
    commentstyle=\color{ipython_cyan}\ttfamily,
    stringstyle=\color{ipython_red}\ttfamily,
    keepspaces=true,
    showspaces=false,
    showstringspaces=false,
    rulecolor=\color{ipython_frame},
    frameround={t}{t}{t}{t},
    numbers=none,
    numberstyle=\tiny\color{halfgray},
    backgroundcolor=\color{ipython_bg},
    basicstyle=\ttfamily\footnotesize,
    columns=fullflexible,
    keywordstyle=\color{ipython_green}\ttfamily,
}

\def\Planck{\textit{Planck}}

\usepackage{amssymb}
\usepackage{comment}

\newcommand{\vnh}{\hat{\mathbf{n}}}
\newcommand{\nobs}{n_g^{\rm obs}}
\newcommand{\nbar}{\bar{n}_g}
\newcommand{\valpha}{\mbox{\boldmath$\alpha$}}
\newcommand{\Mbar}{\bar{M}}

\newcommand{\vbeta}{\mbox{\boldmath$\beta$}}   
\newcommand{\vMsys}{\mbox{\boldmath$M$}}   
\newcommand{\vepsilon}{\mbox{\boldmath$\epsilon$}}

\usepackage{graphicx}	% Including figure files
\usepackage{amsmath}	% Advanced maths commands
 \usepackage{amssymb}	% Extra maths symbols
\usepackage{bm}  % for bold math symbols

\def\apj{{ApJ}}                 % Astrophysical Journal
\def\apjl{{ApJ}}                % Astrophysical Journal, Letters
\def\apjs{{ApJS}}               % Astrophysical Journal, Supplement
\def\ao{{Appl.~Opt.}}           % Applied Optics
\def\aap{{A\&A}}                % Astronomy and Astrophysics
\def\jcap{{J. Cosmology Astropart. Phys.}}
\def\mnras{MNRAS}             % Monthly Notices of the RAS
\def\prd{{Phys.~Rev.~D}}        % Physical Review D
\def\pasp{{PASP}}               % Publications of the ASP
\date{\today}

\begin{document}
\journalinfo{The Open Journal of Astrophysics}
\submitted{submitted December 2024; accepted XXXX}

\shorttitle{J-PLUS. DR3 tomographic analysis of ADF and ARF}
\shortauthors{Hernández-Monteagudo \& J-PLUS collaboration}
\title{J-PLUS: Tomographic analysis of galaxy angular density and redshift fluctuations in Data Release 3. \\
Constraints on photo-$z$ errors, linear bias, and peculiar velocities}

\author{C.~Hern\'andez-Monteagudo$^{\star 1,2}$,
 A.~Balaguera-Antolínez$^{1,2}$,
 R.~von~Marttens$^{9}$,
 A.~del~Pino$^{16,3}$
 A.~Hernán-Caballero$^{3,4}$,
 L.~R.~Abramo$^{5}$,
 J.~Chaves-Montero$^{17}$,
 C.~López-Sanjuan$^{3,4}$,
 V.~Marra$^{6,7,8}$,
 E.~Tempel$^{12}$,
 G.~Aricò$^{18}$,
 J.~Cenarro$^{3,4}$, 
 D.~Cristóbal-Hornillos$^{3}$,
 A.~Marín-Franch$^{3,4}$,
 M.~Moles$^{3}$,
 J.~Varela$^{3}$,
 H.~Vázquez Ramió$^{3,4}$,
 J.~Alcaniz$^{11}$,
 R.~Dupke$^{11}$,
 A.~Ederoclite$^{3,4}$,
 L.~Sodré~Jr.$^{13}$, 
\and R.~E.~Angulo$^{14,15}$}
\thanks{$^\star$ E-mail: \nolinkurl{chm@iac.es} }

\affiliation{$^1$ Instituto de Astrofísica de Canarias (IAC), 
              C/ Vía Láctea, S/N, E-38205, San Cristóbal de La Laguna, Tenerife, Spain} 
\affiliation{$^2$ Departamento de Astrofísica, Universidad de La Laguna (ULL), Avenida Francisco Sánchez,
             E-38206, San Cristóbal de La Laguna, Tenerife, Spain}
\affiliation{$^3$ Centro de Estudios de Física del Cosmos de Aragón,
            Plaza San Juan, 1, planta 3, E-44001, Teruel, Spain}
\affiliation{$^4$ Unidad Asociada CEFCA-IAA, CEFCA, Unidad Asociada al CSIC por el IAA, 
            Plaza San Juan 1, 44001 Teruel, Spain.}
\affiliation{$^5$ Departamento de Física Matemática, Instituto de Física, Universidade de São Paulo, R. do Matão 1371, 05508-090, São Paulo, SP, Brazil}
\affiliation{$^6$  Departamento de Física, Universidade Federal do Espírito Santo, 29075-910, Vitória, ES, Brazil}
\affiliation{$^7$ INAF -- Osservatorio Astronomico di Trieste, via Tiepolo 11, 34131 Trieste, Italy}
\affiliation{$^8$ IFPU -- Institute for Fundamental Physics of the Universe, via Beirut 2, 34151, Trieste, Italy}
\affiliation{$^9$ Instituto de Física, Universidade Federal da Bahia, 40210-340, Salvador-BA, Brazil}
\affiliation{$^{10}$ PPGCosmo, Universidade Federal do Espírito Santo, 29075-910, Vitória, ES, Brazil}
\affiliation{$^{11}$ Observatório Nacional, Rua General José Cristino 77, Rio de Janeiro, RJ, 20921-400, Brazil}
\affiliation{$^{12}$ Tartu Observatory, University of Tartu, Observatooriumi 1, Tõravere 61602, Estonia}
\affiliation{$^{13}$ Instituto de Astronomia, Geofísica e Ciências Atmosféricas, Universidade de São Paulo, 05508-090, São Paulo, Brazil}
\affiliation{$^{14}$ Donostia International Physics Centre (DIPC), Paseo Manuel de Lardizabal 4, 20018, Donostia-San Sebastián, Spain}
\affiliation{$^{15}$ IKERBASQUE, Basque Foundation for Science, 48013, Bilbao, Spain}
\affiliation{$^{16}$ Instituto de Astrofísica de Andalucía, IAA-CSIC, Glorieta de la Astronomía s/n, 18008, Granada, Spain}
\affiliation{$^{17}$  Institut de Física d'Altes Energies, The Barcelona Institute of Science and Technology, Campus UAB, E-08193, Bellaterra, Barcelona, Spain}
\affiliation{$^{18}$ Institut für Astrophysik (DAP), Universität Zürich, Winterthurerstrasse 190, 8057, Zurich, Switzerland\\}

\begin{abstract}
The {\it Javalambre Photometric Local Universe Survey} (J-PLUS) is a {\it spectro-photometric} survey covering about 3,000~deg$^2$ in its third data release (DR3), and containing about 300,000 galaxies with high quality ({\it odds}$>0.8$) photometric redshifts (hereafter photo-$z$s). We use this galaxy sample to conduct a tomographic study of the counts and redshift angular fluctuations under Gaussian shells sampling the redshift range $z\in[0.05,0.25]$. We confront the angular power spectra of these observables measured under shells centered on 11 different redshifts with theoretical expectations derived from a linear Boltzmann code ({\tt ARFCAMB}). Overall we find that J-PLUS DR3 data are well reproduced by our linear, simplistic model. We obtain that counts (or density) angular fluctuations (hereafter ADF) are very sensitive to the linear galaxy bias $b_g(z)$, although weakly sensitive to radial peculiar velocities of the galaxy field, while suffering from systematics residuals for $z>0.15$. Angular redshift fluctuations (ARF), instead, show higher sensitivity to radial peculiar velocities and also higher sensitivity to the average uncertainty in photo-$z$s ($\sigma_{\rm Err}$), with no obvious impact from systematics. For $z<0.15$ both ADF and ARF agree on measuring a monotonically increasing linear bias varying from $b_g(z=0.05)\simeq 0.9\pm 0.06$ up to $b_g(z=0.15)\simeq 1.5\pm 0.05$, while, by first time, providing consistent measurements of $\sigma_{\rm Err}(z)\sim 0.014$ that are $\sim 40~\%$ higher than estimates from the photo-$z$ code {\tt LePhare}, ($\sigma_{\rm Err}^{\rm LePhare}=0.010$). As expected, this photo-$z$ uncertainty level prevents the detection of radial peculiar velocities in the modest volume sampled by J-PLUS DR3, although prospects for larger galaxy surveys of similar (and higher) photo-$z$ precision are promising.
\end{abstract}

\maketitle
%
%-------------------------------------------------------------------

\section{Introduction}
\label{sec:intro}

Physical cosmology is currently entering a singular era where the great advances achieved in the last two decades are going to be critically tested with the advent of the last generation of cosmological surveys of different nature. These experiments will either confirm the current cosmological paradigm (and possibly provide answers for some remaining questions, while probably giving rise to new ones), or will provide definite evidence for the need of new physics in our understanding of the universe. 

From the side of the study of the Large Scale Structure (LSS) of the universe, efforts like {\it Euclid} \citep{euclid_mission}, DESI \citep{desi_survey}, J-PAS \citep{BenitezJPAS,miniJPAS_survey}, SPHEREx \citep{sphereX_survey}, Roman Telescope \citep{roman_telescope}, 4most \citep{4most_survey}, LSST \citep{lsst_survey}, BINGO \citep{Bingo_I}, SKA \citep{ska_survey} are going to mine further and deeper the light and matter distribution of the universe up to $z\sim 5$ (SKA should be able to dig well into the epoch of reionization $-z\sim 10-$ in the next decade). At the same time, experiments measuring anisotropies of the Cosmic Microwave Background (CMB) radiation such as the Simons Observatory \citep[SO,][]{simons_obs}, Stage IV \citep[S4,][]{stage4_cmb}, or LiteBird \citep{litebird_mission} will map with exquisite precision the lensing of both the intensity and polarization anisotropies of the CMB. While the B-mode of the CMB polarization will constrain {\it per se} the inflationary epoch, the gravitational deflection of CMB photons will provide invaluable information on the matter distribution in the visible universe, and on the amplitude of their spatial perturbations on different scales. From these it will be possible to set constraints on the nature of relativistic species (with neutrinos among them) and dark matter \citep{review_DE_taskforce,baumann_18,dvorkin_22}. Furthermore, the combination of LSS and CMB observations will allow accessing physical effects impacting the CMB photon distribution, enabling consistency tests on the cosmological constraints obtained from either the CMB or the LSS side. 

Such consistency is already under exhaustive scrutiny provided the apparent tension on the Hubble constant, $H_0$, as it is measured using the relatively simple, linear physics describing the universe at $z\simeq 1,100$ \citep[$H_0\simeq 68$~km~s$^{-1}$~Mpc$^{-1}$,][to quote just a few]{planck_2018_cosmo_param,act_dr6,boos_eboss_cosmo_param_21,schoneberg_22_BBN_BAO_H0}, or by using standard candles and photometric local distance calibrators \citep[$H_0\simeq 73$~km~s$^{-1}$~Mpc$^{-1}$,][]{riess_shoes_22,Muramaki_noSNIa_inH0_23,Galbany_IR_in_H0_23,Kenworthy_2rung_H0_22,riess_gaia_21}. The tension between these two sets of measurements lies currently at the $5-7~\sigma$ level, depending on the particular data set under analysis \citep[see, for a deep and recent review, ][]{H0_tension_review_Verde}. While it seems hard to reconcile those $H_0$ measurements by modifying the physics of the early universe without ruining other cosmological constraints \citep{knox_millea_20,kamion_riess_22}, latest tests on the local distance calibrators with James Webb Space Telescope (JWST) seem to confirm previous results based upon supernova Type Ia as standard candles and Cepheid stars as local distance calibrators \citep{riess-jwst} \footnote{Very preliminary results by \citet{wfreedman_on_jwst_cchp} suggest that the tension may be associated to the use of the cepheids as distance calibrators, since the use of other calibrating stars (at the tip of the red giant branch or A/B giants in the $J$ region) seem to point to values that are compatible from the early, young universe. To some extent this is however refuted by \citet{riess_on_jwst_cchp_24}.} Future data should then help clarifying the origin of this apparent inconsistency.

In our current, so-called ``concordance" cosmological model \citep[$\Lambda$CDM, see, e.g.,][]{planck_2018_cosmo_param}, most of the energy ($\sim 70~$\%) is in the form of a repulsive force (dubbed as ``dark energy") which, in its simplest form may reduce to a cosmological constant ($\Lambda$) in Einstein's relativistic description of an homogeneous and isotropic universe. About $\sim 5/6$ parts of the remaining energy content is carried by an unknown, invisible type of matter (dark matter), and hence only $\sim 1/6$ of the matter content of the universe (and roughly $\sim 5~$\% of its total energy budget) corresponds to the baryonic matter we are familiar to. Thus, well beyond the tension on $H_0$ (or other parameters of the model) that may or may not survive the upcoming data, the nature of the dominant components of the universe (dark matter and dark energy) remains largely unknown, and this constitutes a formidable task for cosmology and fundamental physics.

In this context of careful, blind and consistent analysis of present and upcoming cosmological datasets, new ideas and approaches to extract cosmological constraints out of the data are key towards a more precise (and accurate) data exploitation. In the last few years a new cosmological observable, named as ``angular redshift fluctuations" (ARF), has been introduced as a new approach to constrain the cosmological density and radial peculiar velocity fields \citep{arf_letter1}. The ARF use the  redshift anisotropies of matter probes selected under any given redshift shell as a complementary observable to the number of matter probes under that same redshift shell (the latter quantity being the standard 2D clustering, hereafter denoted by ``angular density fluctuations" or ADF).  Previous Fisher forecasts \citep{Legrand_ARF} have demonstrated that the addition of ARF on top of ADF and CMB lensing roughly halves the uncertainty on cosmological parameters like $\Omega_m$, $\Omega_b$, $H_0$, or $n_s$, and shrinks the area of the error ellipse by one order of magnitude for the dark energy parameters $(w_0,w_a)$. Roughly at the same time, the implementation of an ARF tomographic analysis on the SDSS DR13 spectroscopic sample of Luminous Red Galaxies \citep[LRGs, ][]{arf_letter2} was able to provide bias measurements at $\sim 5~$\% precision {\em per redshift bin}, and an $\simeq 7$~\% overall measurement of the growth rate, thus yielding among the strongest constraints on deviations from General Relativity (GR), $\gamma = 0.44^{+0.09}_{-0.07}$. ARF have also provided, in a cross-correlation analysis with CMB intensity anisotropies, among the strongest constraints on the kinetic Sunyaev-Zeldovich effect \citep[kSZ,][]{kSZ_sunyaev_80} when using spectroscopic data throughout the redshift range $z\in [0,5]$, \citep{jonas_kSZ_ARF}. 

So far the tightest cosmological constraints from spectroscopic galaxy surveys have been obtained in 3D analysis where a fiducial cosmology is adopted to convert observed quantities (redshifts and angular position on the sky) into 3D cartesian coordinates \citep[see, e.g.,][]{boos_eboss_cosmo_param_21,desi_bao_2024_I,desi_bao_2024_II}. This approach ensures a full utilization of all $k$-modes present in the data, with the fiducial model assumed being actually corrected by accounting for peculiar velocities and imposing statistical isotropy. However, due to curvature effects the interpretation of 2-point statistics on the largest scales is not straightforward, and there remains an implicit assumption of no cosmological evolution inside the 3D box under analysis. Instead, the 2D angular, tomographic approach proposed in, e.g., \citet{balaguera2mass, arf_letter1} adopts narrow redshift shells ($\sigma_z\sim 0.01-0.05$) whose widths are negligible compared to the local Hubble time, and where exact predictions (at least within linear theory) can be made on the largest accessible scales. Of course, any projection under a redshift shell involves the smoothing of some radial modes whose information is lost to some extent, and the combined analysis of a large number of redshift shells may be computationally challenging. Yet consistent results should be obtained via either approach, and each of those may have different limitations and suffer differently from systematics. 

In this work we explore the implementation of a tomographic, angular ADF+ARF analysis of a pseudo-spectroscopic survey like J-PLUS \citep{Cenarro-J-PLUS}. By pseudo-spectroscopic surveys we refer to multi narrow/medium-width optical band surveys like COMBO \citep{combo17_survey}, ALHAMBRA \citep{alhambra_survey}, SHARDS \citep{shards_survey}, PAU \citep{pau_survey}, J-PLUS \citep{Cenarro-J-PLUS}, or J-PAS \citep{BenitezJPAS}, for which, in every pixel in the footprint, there exists a pseudo-spectrum with as many data points as the number of filters carried in the optical system. Typically, for these surveys the spectral resolution roughly equals the number of optical filters, $R\sim N_{\rm filter}$, with $N_{\rm filter}=17, 20, 12$, and $56$ for COMBO, ALHAMBRA, J-PLUS, and J-PAS, respectively. Such modest level of effective spectral resolution can not be compared to typical spectroscopic surveys (for which $R\gtrsim 500$), although it still leaves room for a wealth of science cases provided the photometric redshift precision can easily reach (and surpass) the 1~\% level for a significant fraction of the detected galaxies. For instance,  J-PAS with $N_{\rm filter}=56$ should still be sensitive to the radial BAO \citep{jonas_BAO_photo-zs}, or should provide a superb catalogue of galaxy clusters and groups \citep{begogna_forecasts,lia_clusters,maturi_minijpas}. 

This work formally addresses the use of ADF and ARF with photometric redshifts in real data. This requires implementing the error distribution of those redshift estimates in the predictions for the ADF and ARF angular power spectrum. As we show below, this results in ADF and ARF actually constrainig the typical uncertainties of the photometric redshifts of the galaxy samples under use. Since J-PLUS is a local, relatively shallow survey, our linear theory model for the ADF/ARF angular power spectra will be tested in a non-linear regime, and this analysis will show how ADF and ARF suffer from non-linearities, and how those bias estimates of the galaxies' bias and peculiar velocities. 

In Sect.~\ref{sec:jplusdr3} we introduce and describe the J-PLUS survey, together with its third data release, its photo-$z$ estimation, and the selection of the galaxy sample to analyze. In Sect.~\ref{sec:tomo} we outline our approach to correct for observational systematics, and its outcome. In Sect.~\ref{sec:theory} we describe out theoretical model for the ADF and ARF angular power spectrum, and the parameters we attempt to constrain from their comparison to the actual power spectra measured in J-PLUS DR3 data. In Sect.~\ref{sec:results} we present our results for the ADF-only, ARF-only, and the combined ADF$+$ARF analyses, with some further cross-correlation study of J-PLUS data with {\it Planck}'s CMB convergence $\kappa$ map. We discuss our findings in Sect.~\ref{sec:discussion}, and conclude in Sect.~\ref{sec:conclusions}.

In this work the adopt as fiducial cosmology a flat $\Lambda$CDM model compliant with {\it Planck} 2018 analyses \citep{planck_2018_cosmo_param}, with $\Omega_c h^2=0.122$, $\Omega_b h^2=0.022$ for the dark matter and baryonic matter physical critical density parameters, respectively, $n_S=0.96$ for the scalar spectral index, $h=0.675$ for the reduced Hubble constant, and $A_s=2\times 10^{-9}$ for the amplitude of the scalar curvature power spectrum. Any cosmology-dependent quantity will be computed under this model. 

\section{J-PLUS DR3}
\label{sec:jplusdr3}

In this work we analyse the third data release of the Javalambre Photometric Local Universe Survey (J-PLUS, \url{https://www.j-plus.es}), covering 3,192~deg$^2$ and gathering $\sim$47.4~million objects brighter than $r_{\rm SDSS}<21$ in the dual catalogue (containing entries in at least the $r_{\rm SDSS}$-detection band). As first introduced in \citet{Cenarro-J-PLUS}, J-PLUS is being conducted at the Observatorio Astrofísico de Javalambre \citep[OAJ\footnote{\url{https://oajweb.cefca.es/}},][]{oaj}, at 1,957~m above sea level in the Sierra de Javalambre, Teruel, (Spain). Out of the two main survey telescopes at the OAJ, JAST80 is responsible for J-PLUS: it consists of a M1 84~cm diameter reflective telescope in a German-equatorial mount, whose optical system provides a field of view (FoV) diameter of 2~deg, completely covering the 2~deg$^2$ occupied by the 9.2~kpix $\times$ 9.2~kpix CCD placed at its focal plane. This yields an angular resolution of 0.55~arcmin~pixel$^{-1}$. Besides this CCD, the instrument at its focal plane, T80Cam \citep{t80cam}, carries 12 different optical filters, the standard SDSS {\it g,r,i} and {\it z} broad band filters, plus 8 additional, narrow and medium width ($\sim 200$--$400$~nm) filters placed upon singular stellar spectral features: {\it J0378} sensitive to the [OII]$\lambda3727$ emission line feature, {\it J0395} to the Ca H+K complex, {\it J410} to H$\delta$, {\it J0430} to the $G$ band, {\it J0515} to the Mg~$b$ triplet, {\it J0660} to H$\alpha$, and {\it J0861} to the Ca triplet. Among the latter group we also include the {\it u} filter which differs significantly from its SDSS counterpart. Exposures in each of the filters were set such that typical limiting magnitudes lie in the 20.5--21.5~mag range, and the relative calibration of those bands lies at the mmag level \citep{clsj_jplus_calib}. More details on the telescope, its optical system, and T80Cam can be found in \citet{Cenarro-J-PLUS}.

For every entry in the source catalog, J-PLUS provides an {\it pseudo-}spectrum with 12 entries, each corresponding to each of the optical filters it mounts. Whistle J-PLUS is originally conceived to characterize the stellar populations in the local universe (including our own Galaxy), it has proved to be a superb tool to isolate emission-line objects like star forming galaxies or quasars \citep{spinoso_QSOs,Lumbreras_Calle}. A probability density distribution (PDF) for the photometric redshift (photo-$z$ hereafter) is provided for 44.4~million objects, regardless they are classified as galaxy, quasar, or star. An estimate of the quality of those photo-$z$ PDFs is given by the {\it odds} parameter, defined as \citep{AntonioHC-photozminiJPAS}:
\begin{equation}
odds = \int_{z_{\rm best}-0.03(1+z_{\rm best})}^{z_{\rm best}+0.03(1+z_{\rm best})} dz\,{\cal P}(z|G),
\label{eq:odds}
\end{equation}
where $z_{\rm ml}$ is the redshift maximizing the posterior photo-$z$ PDF ${\cal P}(z|G)$ for a given source being a ``galaxy" ($G$). High values of the {\it odds} parameter should hence correspond to ``packed" or ``self-contained" photo-$z$ PDFs around $z_{\rm best}$, or precise determination of the photo-$z$s. Details on the photo-$z$ estimation follow next.

%-------------------------

\subsection{Photometric redshifts in J-PLUS DR3}

The photo-$z$ of J-PLUS DR3 were obtained using {\sc jphotoz}, a python package that is part of {\sc jype}, the data reduction pipeline for J-PLUS and J-PAS \citep{hectorinprep}. This pipeline acts as an interface between the J-PLUS database and the actual photo-$z$ engine(s) \citep[which in the case of J-PLUS DR3 is {\sc LePhare},][]{Arnouts11}. {\sc jphotoz} handles all the pre-processing of the data that is fed to {\sc LePhare} and the post-processing of its output for ingestion into the database.

The procedure for obtaining photo-$z$ for J-PLUS DR3 with {\sc jphotoz}/{\sc LePhare} is similar to that used for the miniJPAS survey, which is described in detail in \citet{AntonioHC-photozminiJPAS}. Here we summarize the main steps and highlight the differences.

The photometry used is the PSF-corrected photometry (PSFCOR) from the dual-mode catalogs. PSFCOR magnitudes are corrected for Galactic extinction and re-calibrated tile by tile using stellar population synthesis models \citep[see Sects. 2 and 3 in][]{AntonioHC-photozminiJPAS}. {\sc LePhare} works by scanning a redshift range from $z_{\rm min}$ to $z_{\rm max}$ in steps of $z_{\rm step}$. For J-PLUS, $z_{\rm min}$ = 0.0, $z_{\rm max}$ = 1.0, and $z_{\rm step}$ = 0.005 were used. A likelihood distribution $\mathcal{L}$($z$) is obtained from the $\chi^2$ of the best-fitting template at each $z$. The template library, {\tt CEFCA\_minijpas}, contains 50 synthetic galaxy spectra generated with {\sc CIGALE} \citep{Boquien19}. A redshift prior $P(z,T|m)$, where $T$ denotes a galaxy spectra template and $m$ refers to $r$-band magnitude, is derived from galaxy counts in the VIMOS VLT Deep Survey \citep[VVDS;][]{LeFevre05}. Such prior is consequently applied to $\mathcal{L}$($z$) to obtain the redshift probability distribution $P(z)$. For further details on the configuration of {\sc LePhare}, the templates, and the redshift prior, see Sect.~4 in \citet{AntonioHC-photozminiJPAS}.

While the $P(z)$ thus obtained is meaningful only for galaxies, it is computed for all J-PLUS sources brighter than $r=22$ regardless of their classification. This allows to compensate for the classification uncertainty in statistical studies. The mode of $P(z)$ is taken as the best point estimate for the redshift of the source, $z_{\rm best}$. The {\it odds} parameter is defined as the integral of $P(z)$ in the range $z_{\rm best} \pm 0.03(1+z_{\rm best})$ and corresponds to the predicted probability of $\vert\Delta z\vert \equiv (\vert z_{\rm best} - z_{\rm true} \vert)/(1 + z_{\rm true})$ being $< 3~\%$.

In miniJPAS, a contrast correction was applied to the raw $P(z)$ to ensure that the fraction of galaxies with $\vert\Delta z\vert < 3~\%$ matches their average $odds$ value in a spectroscopic sample where $z_{\rm true}$ is known. This correction was not applied for J-PLUS DR3 due to the lack of a representative spectroscopic sample. As a consequence, $odds$ values for J-PLUS DR3 can be overconfident.

%-------------------------
\subsection{Selection of galaxy-type objects in the catalogues}
Our goal in this work is to conduct a tomographic analysis of the galaxy number density and redshift fluctuations throughout the typical redshift depth sampled by J-PLUS. We want to use galaxies for this purpose, and we need identifying which J-PLUS DR3 sources are more like to be galaxies, stars, and quasars. At this step, we make use of two different, machine-learning based methods that have recently been applied on J-PLUS DR3. 

The first one is based upon the work of \citet{RvM}, where the TPOT algorithm \citep{TPOT} examines a large number of Machine Learning (ML) based methods that attempt to classify J-PLUS DR3 sources as galaxies, stars, and quasars. Those methods use a large training sample of spectroscopic data collected by different, external surveys (SDSS DR18 with $\sim$660 thousand galaxies, LAMOST DR8 with 1.2 million stars, and {\it Gaia} DR3 with $\sim$ 230 thousand quasars). They use both photometric and morphological information from all J-PLUS DR3 sources in the classification process. The TPOT algorithm surveyed all those different ML-based models of J-PLUS DR3, and concluded that one of them, XGBoost, provides the best outcome in terms of purity and completeness. In this case, we define as ``galaxy" every object whose probability of being a galaxy exceeds the threshold $0.9$.

The second one has been published in \citet{BANNJOS}, and hereafter will be dubbed as BANNJOS. BANNJOS is a machine learning pipeline relying on Bayesian neural networks to provide PDFs for stars, galaxies and quasars. Towards this aim, it makes use of photometric, morphological and astrometric data from the J-PLUS DR3, {\it Gaia} DR3, and CATWISE2020 surveys, on top of a total of 1.2 million spectroscopic sources gathered from SDSS DR18, LAMOST DR9, DESI EDR, and {\it Gaia} DR3. BANNJOS is proved to provide 95~\% accuracy for all objects brighter than $r_{\rm SDSS}<21$, which is precisely the magnitude cut applied in this work. Since this method provides PDFs for being a star, galaxy, and quasar for every catalog entry, in this case we define ``galaxy" as every object whose median of the PDF of being a galaxy exceeds the threshold value $0.9$.

On top of the restriction on $r_{\rm SDSS}<21$\footnote{The $r_{\rm SDSS}$ magnitude used in this cut is corrected by Galactic extinction using the relation $r^{\rm corr}= r - A_x \times\, ${\tt ebv}, where the {\tt ebv} entry in the database provides the color excess $E(B-V)$ and $A_x=2.383$.}, for both galaxy-candidate catalogs we are also imposing two additional cuts: (i) all sources must have positive most-likely photometric redshift ($z_{\rm ml}>0$), and (ii) all {\tt SExtractor} and mask flags must be zero for all filters. The two galaxy-star-quasar classifiers are independent, and in general yield similar but not identical outputs for a generic sub-sample of J-PLUS DR3 sources. Nevertheless, as we shall show below, once we impose our requirements on the {\it odds} parameter, both classifiers yield galaxy samples that are remarkably close.

\subsection{The high {\it odds} galaxy sub-sample}

\begin{figure}
   \centering
   \includegraphics[width=8cm]{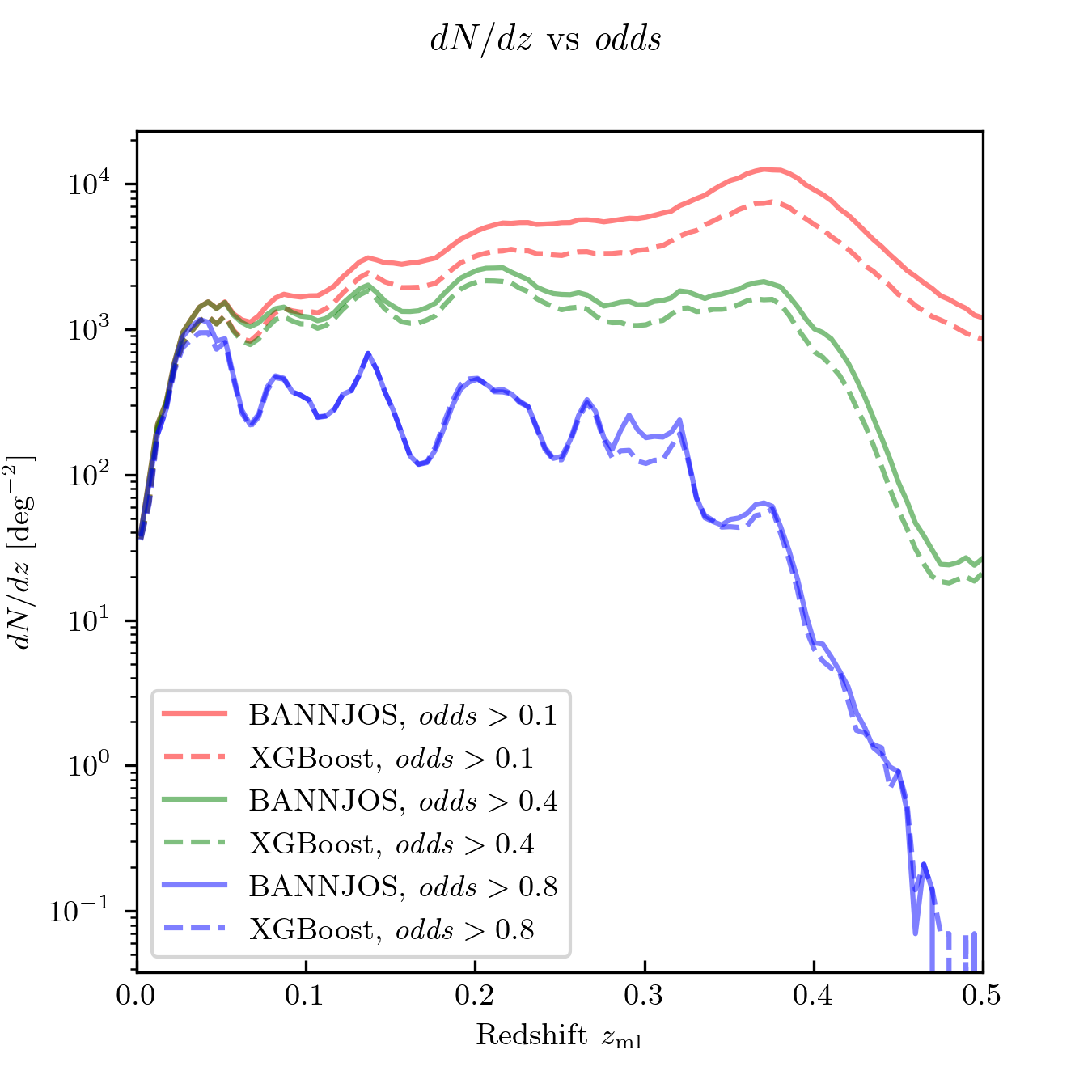}
      \caption{Average angular $dN/dz$ for J-PLUS DR3 $r_{\rm SDSS}<21$ objects labelled as galaxies by the two identification methods for different {\it odds} cuts 
              }
         \label{fig:dNdz_vs_odds}
   \end{figure}

Our initial motivation for considering high {\it odds} galaxies is exploiting the associated high precision photo-$z$ PDFs at searching for signatures of redshift space distortions in their angular density and redshift distributions. However, when considering the two galaxy identification algorithms, it becomes clear that this choice comes with another advantage. As shown in Fig.~\ref{fig:dNdz_vs_odds}, the actual number density for objects seen as ``galaxies" by each of the two algorithms described above is remarkably similar when adopting the threshold $odds > 0.8$ (blue solid and dashed lines in this plot). Given the large number (tens of millions) of galaxy candidates at low $odds$ ($odds > 0.1$), increasing this threshold to $odds > 0.8$ decreases the overall sample to $\sim 300,000$ galaxy candidates, which is still a large enough sample to conduct our tomographic analysis over the $\sim 3,000$ deg$^2$ of J-PLUS DR3. It can also be seen in this plot that the increasing trend of $dN/dz$ with redshift becomes modulated as stricter conditions on the {\it odds} parameter are imposed. 

It must be noted in passing that the shape of the $dN/dz$ plot for either algorithm at high redshifts ($z>0.3$) is remarkably similar to the corresponding curve obtained for those objects labelled as ``stars", where we define as star as every object that, under the cut of $r_{\rm SDSS}<21$, $odds>0.9$, satisfies the condition of {\tt sgcl\_prob\_star}$>0.9$. The latter threshold is applied on the entry for being a star present in the table {\tt jplus.StarGalClass}\footnote{This table is accessible via the Asynchronous ADQL queries link in \url{https://archive.cefca.es/catalogues/jplus-dr3}}. This is a clear indication that stars may leak into the ``galaxy" classification at those high redshift bins poorly probed by the modest size of JAST80 and the limited exposure times of J-PLUS. At the same time, as we shall show below, for redshifts $z>0.15$--$0.2$ we start finding evidence of non-negligible impact of systematics in the angular distribution of our galaxy samples, so most of our analyses focus on the $z<0.2$ redshift range.

%-------------------------------------------------------------
%

%________________________________________________________________
\section{Tomography with maps of ADF and ARF.}
\label{sec:tomo}

The goal of this work is to conduct tomography of the J-PLUS DR3 galaxy angular density and redshift fields. This tomography will be conducted under Gaussian redshift shells of widths $\sigma_z=[0.01,0.03,0.05]$ and nominally centred on a regular redshift array starting at $z_{\rm min}=0.05$ and monotonically increasing with steps of $\Delta z=0.02$ up to $z_{\rm max}=0.39$. While this amounts to 18 different redshifts, we shall see that due to the impact of systematics the actual redshift range under analysis will restrict to $z\in[0.05,0.25]$ or 11 distinct central redshifts, which give rise to $11\times 3=33$ different Gaussian shells after accounting for the three  widths under consideration. We remark that we choose a Gaussian shape for the redshift shells for sake of simplicity, since a priori we could have chosen any other shape. We also emphasize that galaxies belonging to a given redshift shell are selected according to their {\em observed} redshift, which includes errors associated in the photo-$z$ estimation. 

A HEALPix\footnote{HEALPix's URL site \url{http://healpix.sf.net}. Throughout this work we use the python package {\tt healpy}, \citet{healpy}. } \citep{healpix} format sky mask is produced under the resolution parameter $N_{\rm side}=128$, which corresponds to pixels of roughly 0.25~deg$^2$ area. This pixelization is the default used in this work, and allows safely probing harmonic multipoles up to $\ell_{\rm max}=2\times N_{\rm side}=256$. This sky mask is built from the {\tt mangle} format tiles available at \url{https://archive.cefca.es/catalogues/jplus-dr3/}. Initially, after dropping pixels with less than 5~\% of their area covered, the active or useful mask area amounts to 2,859~deg$^2$, but after dropping some few pixels at low galactic latitude with very high star density plus some other region with abnormal depths/exposure times, the effective area for the final mask amounts to 2,795~deg$^2$. This sky mask is fixed for all redshift shells considered in this work.

\subsection{Correction for systematics}

Different observing conditions such as seeing or airmass, in combination with astrophysical factors like star density or Galactic extinction are known to modulate the observed number of galaxies. Therefore, for every redshift shell under study it becomes mandatory to correct for (or at least ameliorate) the impact of the modulation induced by those observational and astrophysical factors. The J-PLUS DR3 database provides access to a number of entries that account for the observing conditions (such as seeing, airmass, limiting magnitude for a fixed aperture, full width-half maximum (FWHM) for every filter, etc), and these can be use to characterize their impact on the observed number of galaxies in every direction/pixel on the sky. 

In this work we shall adopt the approach outlined in \citet{chm-systematics}, which, inspired by \citet{xavier-balaguera19}, attempts to correct for different systematics whose angular pattern on the survey's footprint is known. The model of the observed galaxy angular number density field as a function of the real, underlying galaxy density field and the intervening systematics reads as 

\begin{equation}
\nobs (\vnh) = \biggl( \nbar (1+\delta_g(\vnh)\,) + \valpha \cdot \vMsys(\vnh) \biggr) \prod_{1}^{N_s} (1+\beta_i \delta M_i(\vnh)).
\label{eq:smodel1}
\end{equation}

In this equation $\vnh$ denotes the unit vector pointing to a given direction on the celestial sphere, and each of the $N_s$ potential systematics angular template is broken into their angular mean/monopole and fluctuations,
\begin{equation}
M_i(\vnh) = (\Mbar_i+ \delta M_i (\vnh) ),
\end{equation}
where the index $i$ runs from $1$ to $N_s$, and the amplitude has been re-normalized so that $\langle \delta M_i^2 (\vnh)\rangle_{\vnh}=1$. In Eq.~\ref{eq:smodel1}, $\nbar (1+\delta_g(\vnh)\,)$ denotes the real, underlying galaxy number density field, and the vectors $\valpha$ and $\vbeta$ reflect the additive and mulplicative modulation exerted by the potential systematics template set. In what follows, we introduce $\vepsilon$ via $\valpha = \nbar(1+\vepsilon)$, and accounts for the amplitude of the additive systematics in units of $\nbar$. We shall use $\valpha$ or $\vepsilon$ indistinctly. 

This procedure makes no assumption on the additive or multiplicative nature of those potential systematics, and simultaneously provides estimates (and associated errors) for the $\valpha$ and $\vbeta$ vectors. This method has also been proved to leave minimal residuals (when compared to standard methods that assume either an additive or a multiplicative character for the systematics). We defer the reader to \citet{chm-systematics} for further details on this systematics amelioration algorithm. The set of potential systematics that we consider in the analysis of J-PLUS DR3 is identical to the one presented in that work, and is summarised in their Table~I, that we also incoporate to this section.

\begin{table}
\caption{\label{tab:syst1} Full set of observables upon which we build our potential systematics templates. For a more detailed description of the templates, we refer to \citet{chm-systematics}. }
%\begin{ruledtabular}
\begin{tabular}{ccc}
Index & Observable & Label \\
\hline
0 & Star density & {\tt stars} \\
1 & Zero-point of the image & {\tt zpt} \\
2 & Estimated image noise & {\tt noise}\\
3 & Effective total exposure time  & {\tt teffective} \\
4 & Total exposure time  & {\tt texposed}\\
5 & Number of reduced images combined  & {\tt ncombined} \\
6 & FWHM estimate under Gaussian PSF   & {\tt fwhmg} \\ 
7 & 50\% detection mag. for point-like sources & {\tt m50s} \\
8 & Mag. at SNR=5 and 2$\times$FWHM aperture & {\tt depth2fwhm} \\
9 & Mag. at SNR=5 and 3~arcsec aperture & {\tt depth3as} \\
10 & E(B-V) colour excess from SFD98 & {\tt ebv} \\
11 & E(B-V) colour excess from {\Planck}  & {\tt ebvPlanck} \\
12 & Average airmass of tile & {\tt airmass} \\
13 & Galaxy-weighted average $odds$ parameter & {\tt odds} \\
\hline
 \end{tabular}
%\end{ruledtabular}
\end{table}

%-------------------------------------------------------------
\begin{figure}[ht]
\hspace{0.15cm}
\includegraphics[width=8cm]{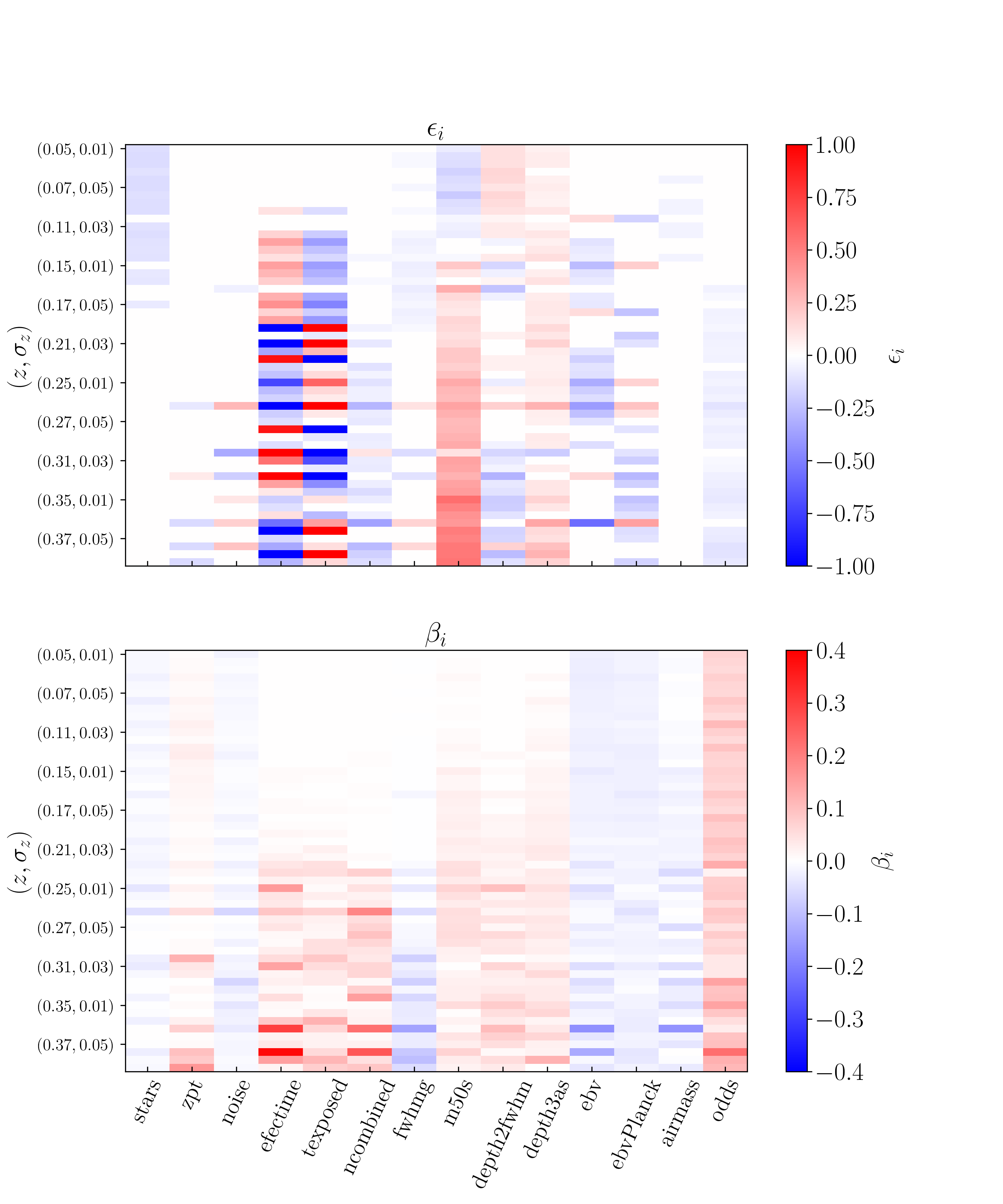}
  \caption{Tomographic systematics correction on J-PLUS DR3 ADF shells for XGBoost galaxy identification method and no assumption about purity.
          }
     \label{fig:Table_syst_APM}
\end{figure}
%-------------------------------------------------------------
%-------------------------------------------------------------
\begin{figure}[ht]
\hspace{0.15cm}
\includegraphics[width=8cm]{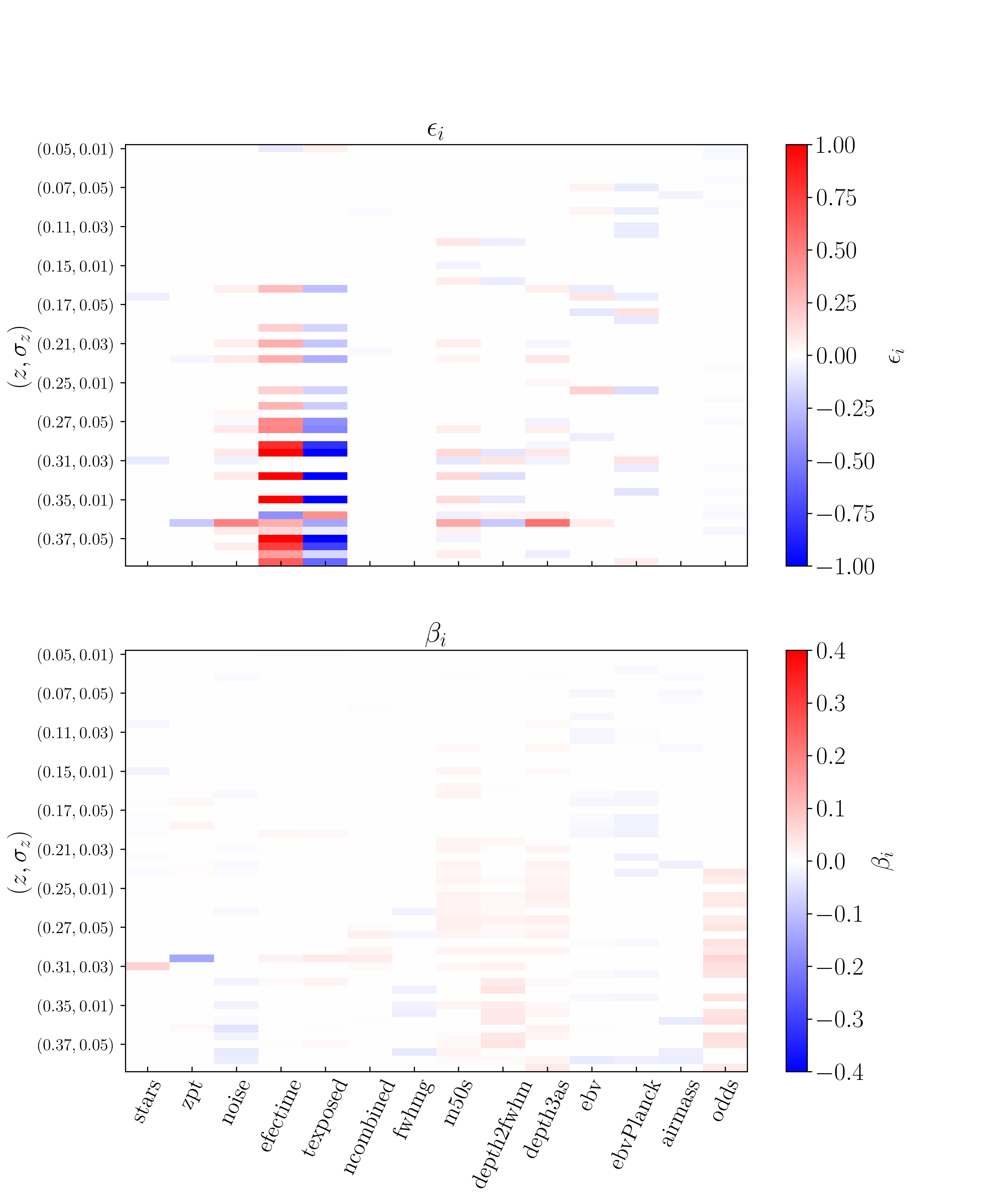}
  \caption{Result of running our systematics amelioration pipeline on our set of already ``processed" J-PLUS-DR3 ADFF maps from the BANNJOS catalogue. See text for details on the non-zero $\epsilon_i$ entries for the {\tt efectime} and {\tt texposed} templates. 
          }
     \label{fig:Table_syst_corr}
\end{figure}
%-------------------------------------------------------------

In the two panels of Fig.~\ref{fig:Table_syst_APM} we display the estimates for $\valpha$ (top panel) and $\vbeta$ (bottom panel) extracted from the BANNJOS galaxy sample. A very similar result is obtained for the XGBoost galaxy sample. Each row corresponds to a combination of $(z_{\rm center}, \sigma_z)$, with $z_{\rm center}$ increasing downwards in both panels, and $\sigma_z$ varying faster than $z_{\rm center}$ along the $Y$-axis. Each column corresponds to a different potential systematics template. Overall, this figure shows that the presence of systematics-induced modulation of the observed galaxy number density increases with redshift, something that is easy to accommodate since the abundance of faint sources are more easily impacted by systematics. The additive coefficients $\epsilon_i$ typically yield larger amplitudes than the multiplicative counterparts, and already for redshifts above $z\sim 0.15$ the $\epsilon_i$ values for several templates like {\tt noise}, {\tt efectime}, {\tt m50s} or {\tt depth2fwhm} depart from zero noticeably. On the other hand, the multiplicative $\beta_i$ amplitudes remain quite small down to $z_{\rm center}\sim 0.25$, with the exception of {\tt stars}, {\tt ebv}, {\tt ebvPlanck}, {\tt airmass}, and {\tt odds} (note the angular correlation of the first three of this template sub-sample). At higher redshifts the $\beta_i$ coefficients for other systematics maps such as {\tt zpt}, {\tt efectime}, {\tt texposed}, {\tt ncombined},  {\tt fwhmg}, take non-zero values.

%-------------------------------------------------------------
   \begin{figure*}
   %\centering
    \hspace*{-2.cm}
   \includegraphics[width=0.55\paperwidth]{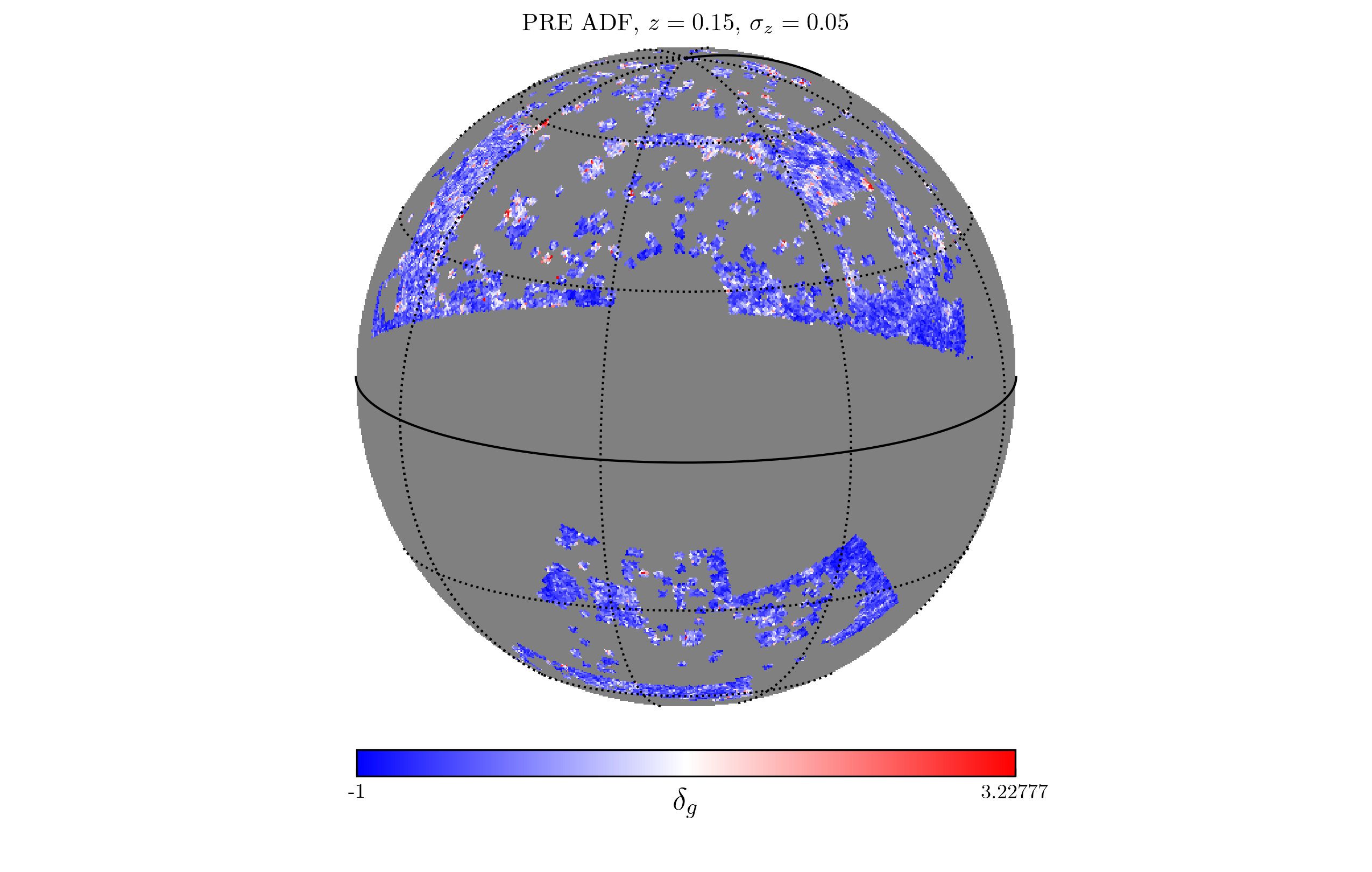}
    \hspace*{-1.cm}
   \includegraphics[width=0.55\paperwidth]{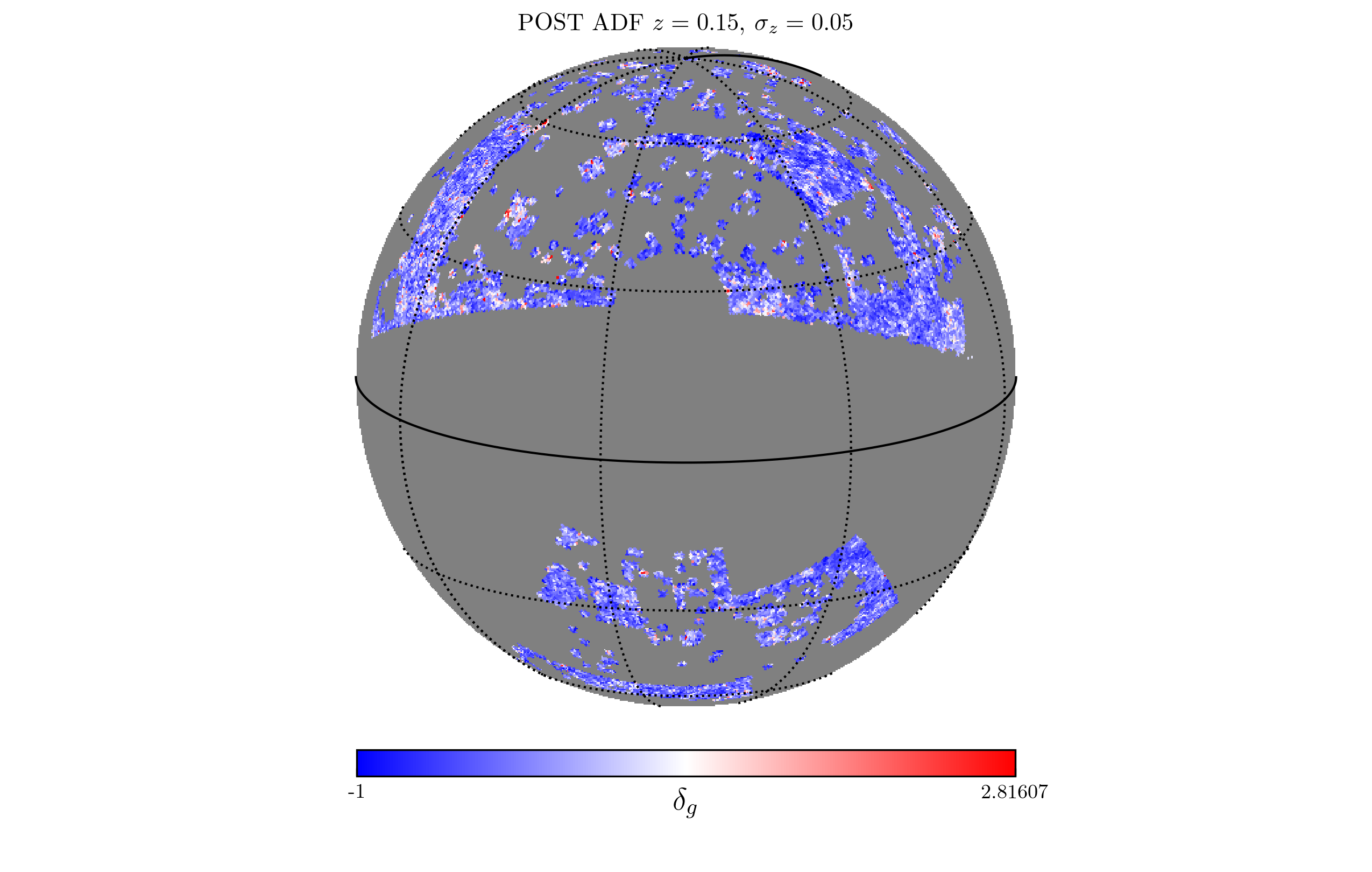}
    \hspace*{-2cm}
   \includegraphics[width=0.55\paperwidth]{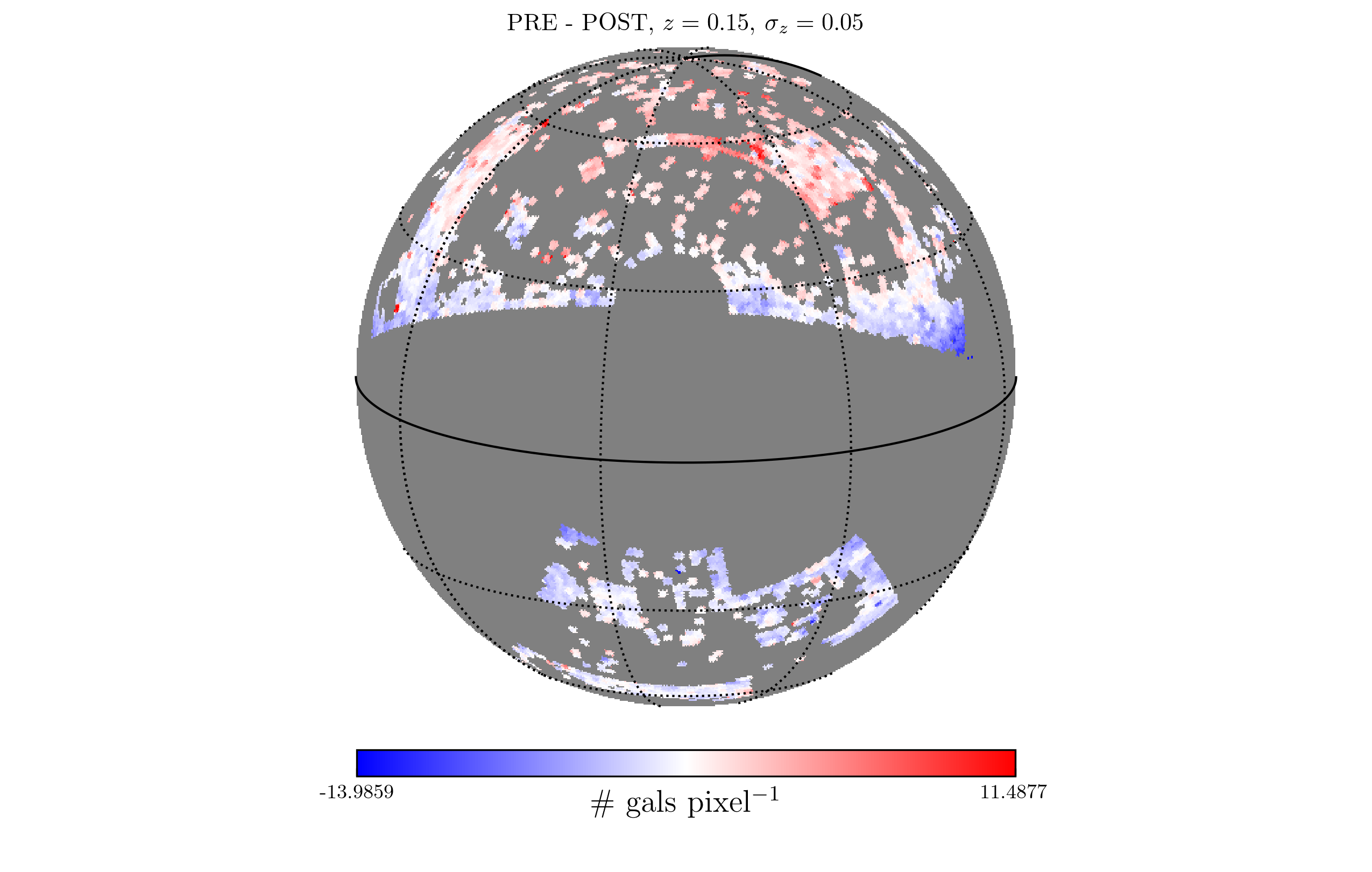}
    \hspace*{-1.cm}
   \includegraphics[width=0.55\paperwidth]{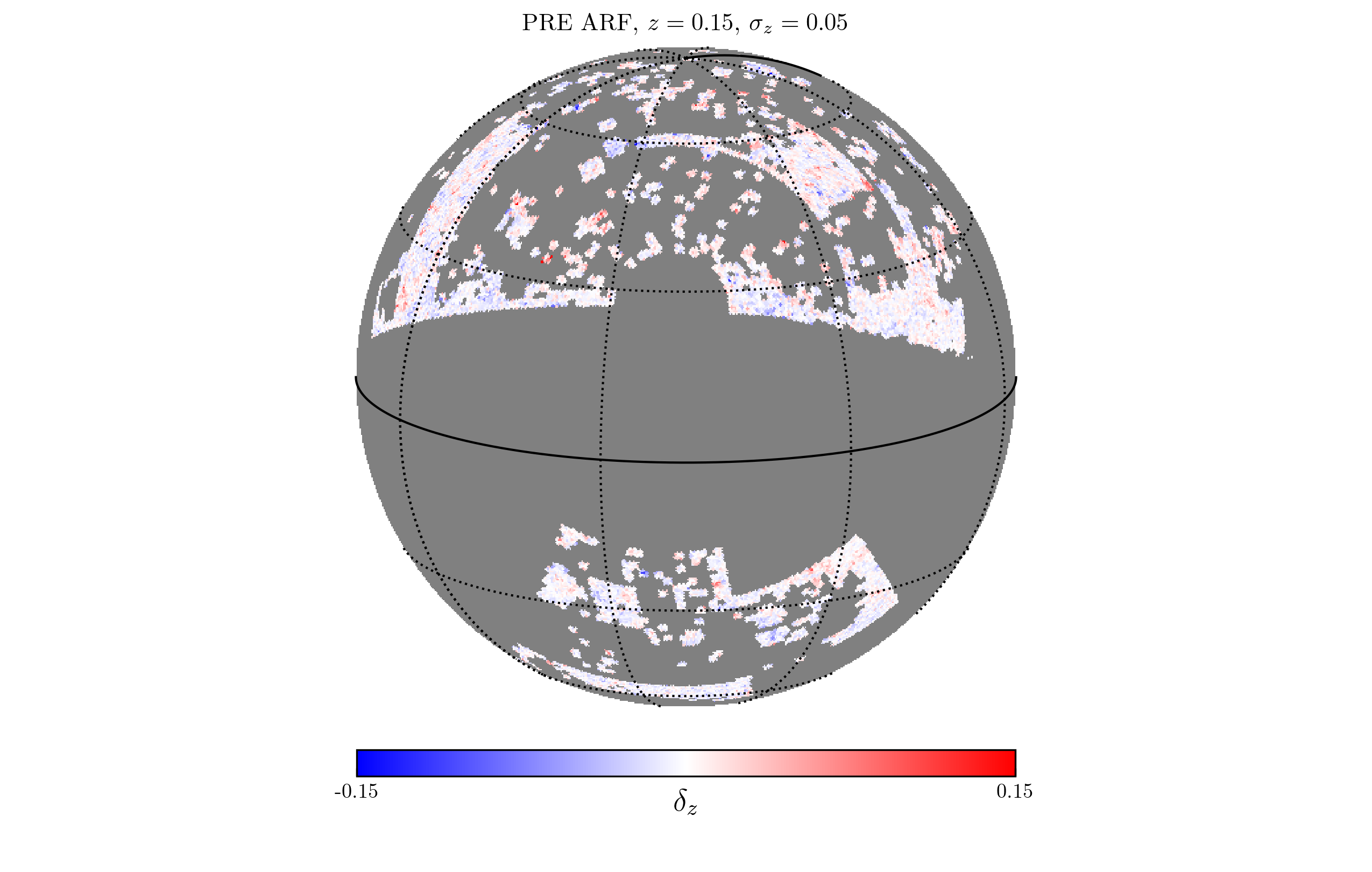}
      \caption{{\it Top, left panel:} Angular density fluctuation map (or ADF/density contrast $\delta_g(\vnh)$ map) from the BANNJOS galaxy sample, computed under the Gaussian shell centered upon $z_{\rm center}=0.15$ and with width $\sigma_z=0.05$ {\em before} correcting for systematics. All maps are given in Galactic coordinates. {\it Top, right panel:} Same as in previous panel, but {\em after} correcting for systematics. {\it Bottom, left panel:} Difference or correction map, that amounts to $n_g^{\rm obs}(\vnh) - n_g^{\rm corr}(\vnh)$, in units of number of galaxies per pixel. {\it Bottom-right panel:} ARF map corresponding to the same redshift shell following the first definition given in Eq.~\ref{eq:arf_defI}.
              }
        
         \label{fig:maps_corr}
   \end{figure*}
%-------------------------------------------------------------

We test the level of residuals remaining in our ``corrected" maps from the BANNJOS catalogue by running again our systematics pipeline on them. The result of this test is shown in Fig.~\ref{fig:Table_syst_corr}, and leaves practically no evidence for systematics residuals with the exception of the {\tt efectime} and {\tt texposed} templates, which yield opposite entries for the $\epsilon_i$ additive parameters in practically every redshift shell. Provided these two templates are extremely similar (and thus highly correlated), it is fair to take the average of their entries, which are very close to zero and carry little to negligible statistical significance. Repeating this same analysis on the ``corrected" maps from the XGBoost catalogue gives no significant hint of residuals, for any template nor redshift shell or width.

In our tomographic analyses, under each Gaussian redshift shell, we consider two different projections on the 2D sphere. The first one constitutes the standard galaxy number density contrast $\delta_g$ (that we dub here as ``angular density fluctuations" or ADF). This observable is computed from a map containing the number of galaxies under the Gaussian shell that fall per pixel as
\begin{equation}
\delta_g(\vnh) = \frac{n_g(\vnh)}{\langle n_g \rangle_{\vnh}}-1,
\label{eq:adf1}
\end{equation}

where $n_g(\vnh)$ is the galaxy number count per pixel, either before or after correcting for systematics. 

We also consider a second observable, the so-called angular redshift fluctuations \citep[hereafter ARF,][]{arf_letter1}, that instead of counting galaxies per pixel, it observes the redshift deviations with respect to the mean redshift under the shell. Depending upon its normalization, there is room for two, slightly different definitions for the ARF:
\begin{enumerate}
\item The first definition refers the redshift fluctuations with respect to average number density of sources:
\begin{equation}
\biggl(\delta z (\vnh) \biggr)^I:= \frac{\sum_{j\in\vnh}W_j (z_j-\bar{z})}{\langle \sum_{j\in\vnh}W_j  \rangle_{\vnh}}.
\label{eq:arf_defI}
\end{equation}
In this equation, the weight $W_j$ provides the Gaussian distance (in redshift) of the $j$-th galaxy falling in pixel $\vnh$ from the central redshift of Gaussian shell: $W_j = \exp{-(z_j-z_{\rm center})^2/(2\sigma_z^2)}$. The redshift $\bar{z}$ refers to the average redshift under the shell, $\bar{z}=\sum_{{\rm all}\,j} W_j z_j / \sum_{{\rm all}\,i} W_i$, and is typically close to $z_{\rm center}$ (depending on the slope of $dN/dz$). The denominator of the equation above contains the angular average (denoted by $\langle ... \rangle_{\vnh}$) of the number of galaxies under the shell.

\item The second definition normalizes rather by the amount of galaxies in {\em each} pixel, rather than by its average. It is thus less stable in case of sparse galaxy surveys with a low number of galaxies per pixel, but at the same time it is less sensitive to systematics \citep{arf_letter1}:
\begin{equation}
\biggl(\delta z (\vnh) \biggr)^{II}:= \frac{\sum_{j\in\vnh}W_j (z_j-\bar{z})}{\sum_{j\in\vnh}W_j }.
\label{eq:arf_defII}
\end{equation}

\end{enumerate}

%-------------------------------------------------------------
   \begin{figure*}
   \centering
   \includegraphics[width=0.8\paperwidth]{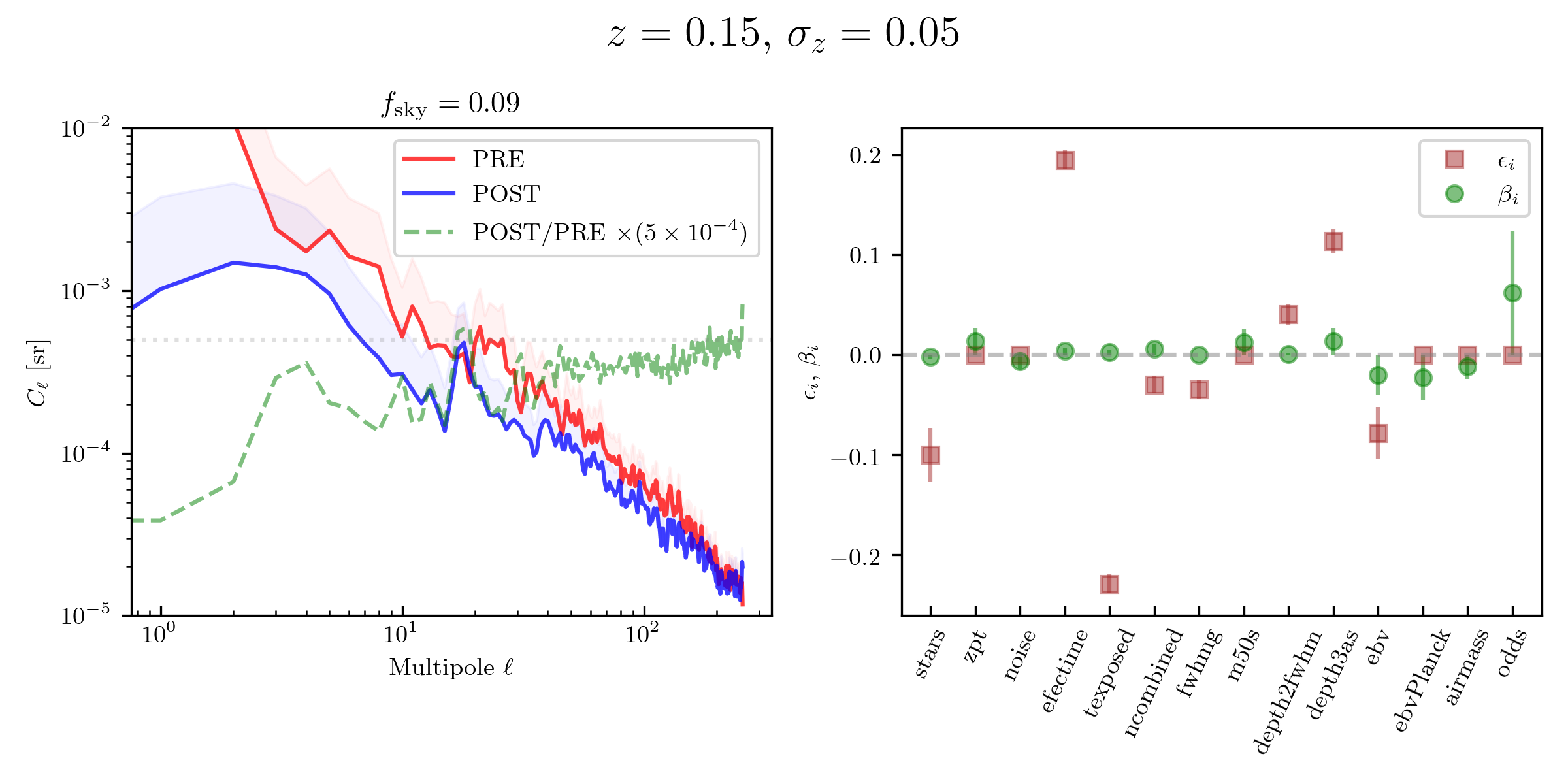}
      \caption{Impact of systematics amelioration on the ADF map at $z_{\rm center}=0.15$ and $\sigma_z=0.05$ computed on top of the BANNJOS galaxy sample. The left panel shows the shot-noise corrected pseudo-angular spectrum (no mask correction applied) before (blue) and after (red) correction for systematics. The ratio of the two spectra, scaled to $5\times 10^{-4}$, is also given by the green, dashed line. The amplitude of the additive ($\epsilon_i$) and multiplicative ($\beta_i$) correction for each systematics template is given in the right panel.
              }
         \label{fig:cls_corr}
    \end{figure*}
%-------------------------------------------------------------

Both definitions, despite being formally different, converge to the same expression at the limit of linear theory of cosmological perturbations when considering only the so called {\tt density} and {\tt redshift}/Doppler terms, \citep{adal_chm}. In what follows we shall stick to the first definition (and we shall refer to it simply as $\delta z (\vnh)$, without super-indexes), unless otherwise explicited. Under this linear theory, both ADF and ARF should be Gaussian fields that are completely determined by their second order momenta, either the angular correlation function $w(\theta)$ or its harmonic transform, the angular power spectrum $C_{\ell}$ for every harmonic multipole $\ell$ (they are related via $w(\theta)=\sum_{\ell}(2\ell+1)/(4\pi) C_{\ell} P_{\ell}(\cos \theta)$, with $P_{\ell}(x)$ the $\ell$-th order Legendre polynomial). Standard Boltzmann codes like {\tt CAMB} \citep{camb_code} or {\tt CLASS} \citep{class_code} provide predictions for the angular power spectrum of the ADF under a large variety of cosmological scenarios. More recently, \citet{adal_chm} modified the code {\tt CAMB} into {\tt ARFCAMB}\footnote{{\tt ARFCAMB} can be accessed at \url{https://github.com/chmATiac/ARFCAMB}} so that it also provides linear theory predictions for ARF. {\tt ARFCAMB} is then our reference code when comparing to theoretical predictions in this work (see Sect.~\ref{sec:theory}).

In Fig.~\ref{fig:maps_corr} we display maps (in Galactic coordinates) of the galaxy ADF or density contrast $\delta_g (\vnh) = n_g(\vnh) / \langle n_g \rangle_{\vnh}-1$ before (left, top panel) and after (right, top panel) correcting for systematics. The bottom, left panel shows the difference or correction map $n_g^{\rm obs}(\vnh)-n_g^{\rm corr}(\vnh)$. It can be seen that the raw map tends to miss galaxies and structure at low galactic latitudes, both in the northern and southern Galactic hemispheres, while a very dense structure at $\delta_{\rm DEC}\approx 60$~deg. is smoothed in the corrected map. Typical ADF map amplitudes range from $\delta_g\simeq -1$ up to $\delta_g \sim 3$. Instead, the ARF map shown in the bottom-right panel of this figure shows a very different pattern: by construction it is symmetric around zero, and it displays a very Gaussian histogram, rarely reaching $+/-\,3\,\sigma$ excursions.  

In Fig.~\ref{fig:cls_corr} we show the outcome of our systematics amelioration scheme. The left panel displays the pseudo-angular power spectrum computed from the pre- (red) and post-correction (blue) ADF map. The correction is significant in practically the entire $\ell$-multipole range (with the exception of the smallest angular scales/high $\ell$-s). The ratio of the corrected angular pseudo-power spectra over the raw one is given by the green, dashed line: this ratio is normalized to $5\times 10^{-4}$ for visualization purposes, reflects larger correction on larger angular scales (lower $\ell$s).  The right panel, instead, provides the actual estimates for $\valpha$ and $\vbeta$. Some of $\epsilon_i$ values are at the $\sim 0.1$--$0.2$ level, which is not negligible. Note that, given the strong correlated pattern between {\tt efectime} and {\tt texposed}, the corrections associated to these two templates partially cancel each other. For redshifts deeper than $z_{\rm center}=0.15$, the impact of systematics are typically more important than shown in this plot.

A particular case is the one associated to the {\tt odds} template, which, according to the rightmost columns in Fig.~\ref{fig:Table_syst_APM}, presents consistent positive values for multiplicative factor $\beta_i$ at the level of $10~\%$ for practically all redshift shells. This potential source of systematics was carefully addressed in \citet{Marti2013}, who found {\it odds} to be impacting strongly the effective galaxy bias. One must note that the {\it odds} parameter is typically higher for brighter galaxies for which photo-$z$ estimates have higher quality. As a result high-{\it odds} galaxies reside in high mass halos whose bias is typically higher than a generic galaxy population at the same redshift. While in the case of \citet{Marti2013} the impact of the {\it odds} parameter was corrected at the level of the 2-point statistics, in our case we are attempting its correction at the {\em map} level. But in either case the following concern arises: provided that the {\it odds} parameter is correlated to a physical property of the galaxies (luminosity or mass), by correcting for it we may be (artificially) modifying the inferred physical bias of our sources. To some extent this in unavoidable provided our basic, underlying assumption about systematics templates is that they are not correlated to the physics of the real galaxy sample, something we expect to (at least partially) fail for the {\it odds} parameter. This must be kept in mind, even if the (possible) modulation of the bias seems rather constant throughout redshift bins, and always at the level of $\sim +10~\%$. 

Unfortunately, it is not trivial to export the corrections on the ADF map derived from our systematics amelioration algorithm to the ARF maps since this would require having knowledge on how those corrections apply {\em under} the Gaussian redshift shell. As already shown in \citet{arf_letter1}, the second ARF definition given in Eq.~\ref{eq:arf_defII} is unsensitive to either additive or multiplicative systematics as long as these remain constant under the redshift shells, not a too demanding requirement provided the adopted shell widths are relatively narrow ($\sigma_z\leq 0.05$). In what follows we will test the robustness of ARF by adopting also Eq.~\ref{eq:arf_defII}, bearing in mind its possible instabilities for low density shells.

\subsection{Correlation matrices}

As suggested by the typical width of the galaxy-galaxy 3D correlation function, most of the clustering of the large scale structure (LSS) shows up at scales of $\sim [5,20]~h^{-1}$~Mpc, while systematics associated to observing conditions and astrophysical foregrounds project along the line of sight on much larger scales. Therefore, a direct test for the presence of residual systematics consists on testing the correlation between different redshift shells lying at very different redshifts. 

If we let $\delta_i (\vnh)$ represent the ADF or ARF map at any redshift shell $i$, then we can decompose it via
\begin{equation}
\delta_i(\vnh) = \sum_{\ell,m} \delta^i_{\ell,m} Y_{\ell,m}(\vnh),
\label{eq:sphharm_decomp}
\end{equation}
where the $Y_{\ell,m}(\vnh)$ constitute the orthonormal basis of spherical harmonics. One can then easily cross-correlate different redshift shells via the cross angular power spectrum
\begin{equation}
C^{i,j}_{\ell} = \frac{\sum_{m=-\ell,\ell} \delta^i_{\ell,m} (\delta^j_{\ell,m})^\star}{(2\ell+1)},
\label{eq:xps}
\end{equation}
with $(\delta^i_{\ell,m})^\star$ denoting the complex conjugate of $\delta^i_{\ell,m}$. This cross power spectrum can again be written as the harmonic transform of the angular cross-correlation function $w^{i,j}(\theta)$:
\begin{equation}
w^{i,j}(\theta) = \langle \delta^i(\vnh_1) \delta^j(\vnh_2)\rangle_{\vnh_1\cdot\vnh_2 = \cos \theta} = 
\frac{2\ell+1}{4\pi} C^{i,j}_{\ell} P_{\ell} (\cos \theta).
\label{eq:xcf}
\end{equation}

In order to quantify the level of systematics induced correlation between distant redshift shells, we compute the cross angular power spectrum between all possible shells, add for all multipoles, and define the following correlation matrix:
\begin{equation}
{\rm Corr}^0_{i,j}=\sum_{\ell=2}^{2\,N_{\rm side}} C^{i,j}_{\ell},
\label{eq:corr_matrix0}
\end{equation}
which we choose to normalize by its diagonal,
\begin{equation}
{\rm Corr}_{i,j} = 
\frac{ {\rm Corr}^0_{i,j} }{\sqrt{{\rm Corr}^0_{i,i}{\rm Corr}^0_{j,j}}}.
\label{eq:corr_matrix1}
\end{equation}

\begin{figure*}
   \centering
    %\hspace*{-1.cm}
   \includegraphics[height=0.3\paperheight]{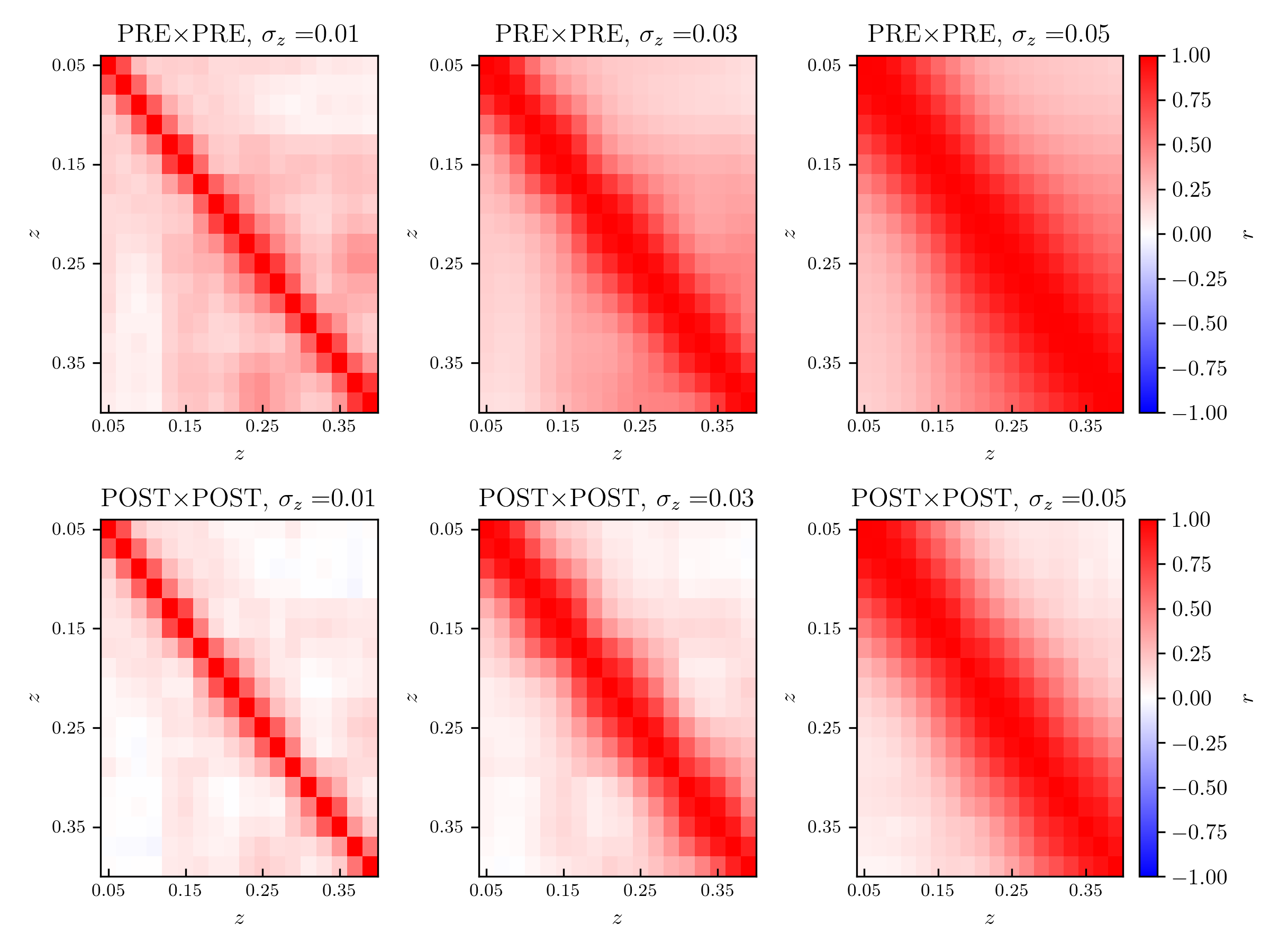}
    %\hspace*{1.cm}
   \includegraphics[height=0.3\paperheight]{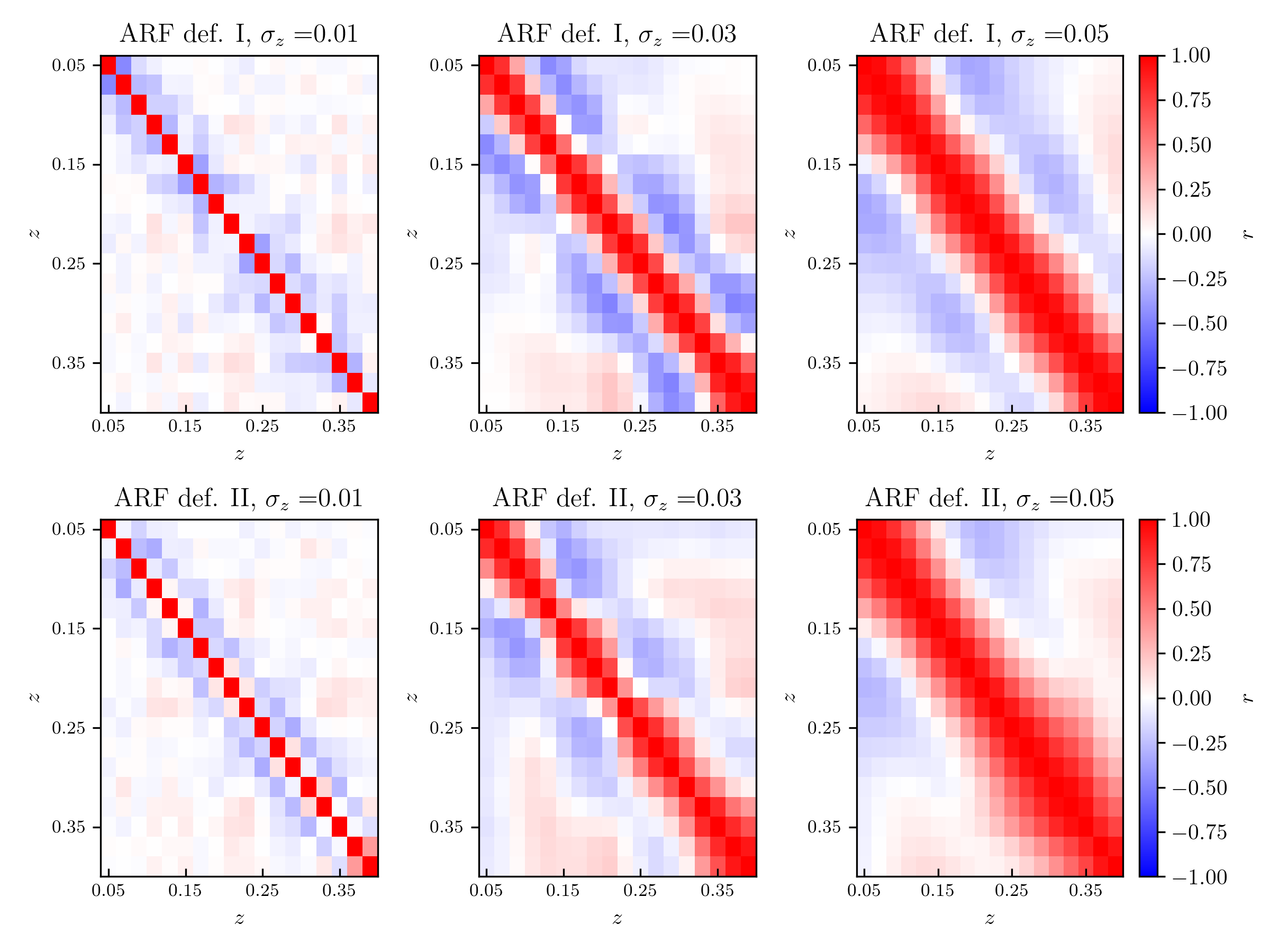}
    \caption{{\it Top panel:} Cross-correlation coefficient matrix for ADF before (top row) and after (bottom row) correction for systematics (BANNJOS galaxy catalogue). A lower level of cross-correlation between distant redshift shells can be seen in the latter case. {\it Bottom panel:} Cross-correlation coefficient matrix for ARF. The matrix at the top row corresponds to definition I of the ARF, whereas the bottom one displays the result for definition II. Both correlation patterns are very similar, with the exception of the highest redshift shells. 
              }
    \label{fig:Table_syst_RvM}
   \end{figure*}

In Fig.~\ref{fig:Table_syst_RvM} we display the cross-correlation matrix ${\rm Corr}_{i,j}$ obtained from ADF and ARF maps built from the XGBoost catalog (being the corresponding matrix for the BANNJOS catalog very similar). The top row in the top panel provides the cross-correlation matrix for the ADF before correcting for systematics, while the bottom row does for the post-correction case. Each column corresponds to a different value of the shell width $\sigma_z$. The off-diagonal structure in the top row is clearly positive, pointing to the presence of systematics spanning through very different redshift shells. This structure disappears to a notable extent in the bottom row, obtained after correcting for systematics: off-diagonal terms are now closer to zero, although an appreciable level of positive entries remain. This positive character points to some systematics residuals (otherwise one would expect a symmetric distribution of off-diagonal elements around zero).

The bottom panel contains a similar exercise conducted on ARF maps. In this case both rows use ARF maps without any systematics correction, the difference lying in the ARF definition adopted. The top row uses the first ARF definition given in Eq.~\ref{eq:arf_defI}, while the bottom row uses Eq.~\ref{eq:arf_defII}. At first sight it is obvious that the ARF correlation properties differ significantly to those of ADF, as first shown by \citet{Legrand_ARF}: the positive diagonal is followed by an anti-correlated set of nearby neighbors that disappears as the redshift distance increases. We can see that, at low redshifts, both matrices display these correlation properties, with far off-diagonal terms oscillating around zero, particularly for the narrowest width (left column). Differences however arise at higher redshifts, for which the anti-correlated band of neighbors tends to vanish in the bottom row (corresponding to the second ARF implementation of Eq.~\ref{eq:arf_defII}). While the reason for this mismatch is unclear, it restricts to the extremely high redshift shells J-PLUS can reach for regular galaxies, for which more aggressive systematics corrections are required, under significantly lower galaxy number densities. 

%-------------------------------------------------------------

\section{A linear theoretical reference model for non-linear scales}
\label{sec:theory}

The J-PLUS DR3 survey is sampling the local universe with a relatively modest depth of $z\sim 0.3$, with most of our analyses restricting to yet lower redshifts ($z_{\rm max}\approx 0.2$) given the non-trivial modulation of systematics. This redshift corresponds to a depth of roughly $\sim 600~h^{-1}$~Mpc, and most of our statistical power will come from scales at least  one order of magnitude smaller, i.e., tens of $h^{-1}$~Mpc. On such scale range cosmological predictions at the linear level of the theory of perturbations are known to be innacurate at best at the 10--30~\% level \citep[see,][for example]{mice_fosalba_15}.

Yet this degree of accuracy is admissible for the purposes of this work. We aim at comparing ADF and ARF (and their combination) as prospective cosmological probes in spectro-photometric surveys like J-PAS. Our goal is studying the redshift evolution of the linear bias, and setting the best possible constraints on the amplitude of the radial peculiar velocities (the so-called redshift-space-distortions term). We are also interested in studying the behavior of ADF and ARF in a non-linear regime and how they compare to linear theory predictions. At the same time, developing corrections beyond linear theory is motivated when cutting-edge cosmological measurements are under reach, which is clearly not the case in this study.

We therefore shall compare our measurements to linear theory prediction provided by the Boltzmann code {\tt ARFCAMB}, that incorporates ARF on top of the standard cosmological observables (like ADF, weak lensing, and intensity, polarization, and lensing angular anisotropies of the Cosmic Microwave Background radiation), and all of their possible cross-correlations.

Boltzmann codes like {\tt CAMB} or {\tt CLASS} include a set of linear theory general-relativistic (GR) corrections when computing the angular power spectra of the ADF. Those GR corrections are also considered in the ARF computation within {\tt ARFCAMB}. These corrections are added on top of the leading contribution given by the {\tt density} term, tracking the density of the number counts of galaxies. Out of all relativistic corrections, the so-called {\tt redshift} correction, accounting for the radial gradient of the galaxies' peculiar velocity is in most cases the one giving the largest contribution (particularly for narrow shells). Next leading correction is typically the {\tt lensing} term, although this is only relevant for wide shells placed at high redshifts (for which the {\tt redshift} term is usually negligible). Other peculiar velocity-related terms like the {\tt radial} term (inversely proportional to the distance to the sources) are only relevant to the largest angular scales, that we can hardly probe with our modest footprint of $\lesssim 3,000$~deg$^2$. 

Hence we shall consider two terms for both ADF and ARF, namely the {\tt density} and {\tt redshift} terms, which actually correspond to the {\it density} and {\it velocity} contributions to the transfer function defined in \citet{arf_letter1}. While the {\it density} term scales with redshift as $b(z)\sigma_8(z)$, the {\it velocity} contribution is proportional to $E(z)f(z)\sigma_8(z)$, with $E(z)=\sqrt{\Omega_m(1+z)^3+\Omega_{\Lambda}}$ in a flat universe, $f(z)=d\log D_\delta/d\log a$ the growth rate given by the logarithmic derivative of the matter density growth factor $D_\delta$, and $\sigma_8(z)$ the linearly extrapolated rms of the matter density contrast averaged inside spheres of comoving radius $8~h^{-1}$~Mpc. Most of the cosmological information is encoded in the growth of density perturbations given by $D_\delta(z)$ and its derivative $f(z)$. From ADF/ARF we can however only access the combinations $b_g(z)\sigma_8(z)$ and $E(z)f(z)\sigma_8(z)$. Since the angular power spectrum is a two-point statistics derived from the ADF/ARF fields, it can be written as the sum of the square of each two terms plus a cross term:
\begin{equation}
C^{\rm theory}_{\ell} = b_g^2\, C_{\ell}^{\delta,\,\delta} + 2\, b_g\, C^{\delta,\,vlos}_{\ell} + C_{\ell}^{vlos,\,vlos}.
\label{eq:celle_theory}
\end{equation}
In this equation, the galaxy linear bias $b_g$ is coupled only to the {\tt density} term, and is a priori redshift dependent, although we assume it to be a constant provided we are considering relatively narrow redshift shells. The modeling observed angular power spectrum differs from the equation above since {\it (i)} the ADF and ARF are sampled by discrete objects (galaxies) and this introduces a sampling (or shot noise) term, and {\it (ii)} due to small sky coverage (the fraction of the sky covered by J-PLUS DR3 is about $f_{\rm sky}\simeq 0.07$) the $C_{\ell}$ estimates are biased and coupled (for different $\ell$s they are no longer independent as under full sky coverage). We thus model the observed angular power spectrum as
\[
C^{\rm obs}_{\ell} = {\cal M}_{\ell,\ell'} \biggl[ b_g^2\, C_{\ell'}^{\delta,\,\delta} + 2\, b_g\,{\cal A}_v C^{\delta,\,vlos}_{\ell'} + {\cal A}_v^2\, C_{\ell'}^{vlos,\,vlos} \biggr] 
\]
\begin{equation}
\phantom{xxxxxxxxxxxxxxx}
+ C^{\rm shot\,noise}_{\ell}.
\label{eq:celle_obs}
\end{equation}

Here, we have introduced a coupling matrix ${\cal M}_{\ell,\ell'}$ that accounts for the bias and the coupling on the raw $C_{\ell}$s from a limited sky coverage (there is a summation on $\ell'$ implicit in this equation). The shot noise term $C_{\ell}^{\rm shot\,noise}$ refers to {\em raw} $C_\ell$s as they are estimated from the masked sky. Finally, given the low value of $f_{\rm sky}$, we choose to bin $C_{\ell}$ into multipole bins of width $\Delta \ell=10$, starting at $\ell_{\rm min}=12$. Binning the multipoles into bins is equivalent to multiplying by a new matrix of dimension ${\cal B}_{L,\ell}=1/\Delta\ell$ if $\ell\in L$ and ${\cal B}_{L,\ell}=0$ otherwise. Our final data model is then given by 
\[
C^{\rm obs}_{L} = {\cal B}_{L,\ell}  
{\cal M}_{\ell,\ell'}
\biggl[ b_g^2\, C_{\ell'}^{\delta,\,\delta} + 2\, b_g\,{\cal A}_v C^{\delta,\,vlos}_{\ell'} + {\cal A}_v^2\, C_{\ell'}^{vlos,\,vlos} \biggr]
\]
\begin{equation}
\phantom{xxxxxxxxxxxxxxxxxxx}  
+ {\cal B}_{L,\ell} C^{\rm shot\,noise}_{\ell}.
\label{eq:celle_binned}
\end{equation}

Equation~\ref{eq:celle_binned} reads as the model vector whose difference with respect to the data vector is to be minimised. {\it A priori}, there exists a $C_\ell$ vector for every redshift shell and every width ($z_{\rm center}, \sigma_z$) for ADF and ARF, so our full data vector considers all those redshift shells. Our model vector depends on the galaxy linear bias, $b_g$, which we render as a separate parameter for every value of $z_{\rm center}$ (although we assume that it does not depend upon $\sigma_z$). In contrast, the ${\cal A}_v$ parameter giving the amplitude of the {\tt redshift} term\footnote{This amplitude ${\cal A}_v$ parameter is taken with respect to the amplitude of our fiducial model, i.e., ${\cal A}_v:= (Ef\sigma_8)/(Ef\sigma_8)_{\rm fid}$.} is assumed constant throughout the entire depth of our sample. The shot-noise term $C_{\ell}^{\rm shot\,noise}$ is computed, for each shell, by keeping the photo-$z$ values of the real galaxy sample and shuffling randomly their angular positions within the sky mask. Our estimate of $C_{\ell}^{\rm shot\,noise}$ is the average of the resulting angular power spectra from such $N_{\rm random}=20$ random assignments of angular coordinates.

\subsection{ADF and ARF sensitivity to photo-$z$ errors}

%________________________________________________________________

   \begin{figure*}
   \centering
   \includegraphics[width=0.4\paperwidth]{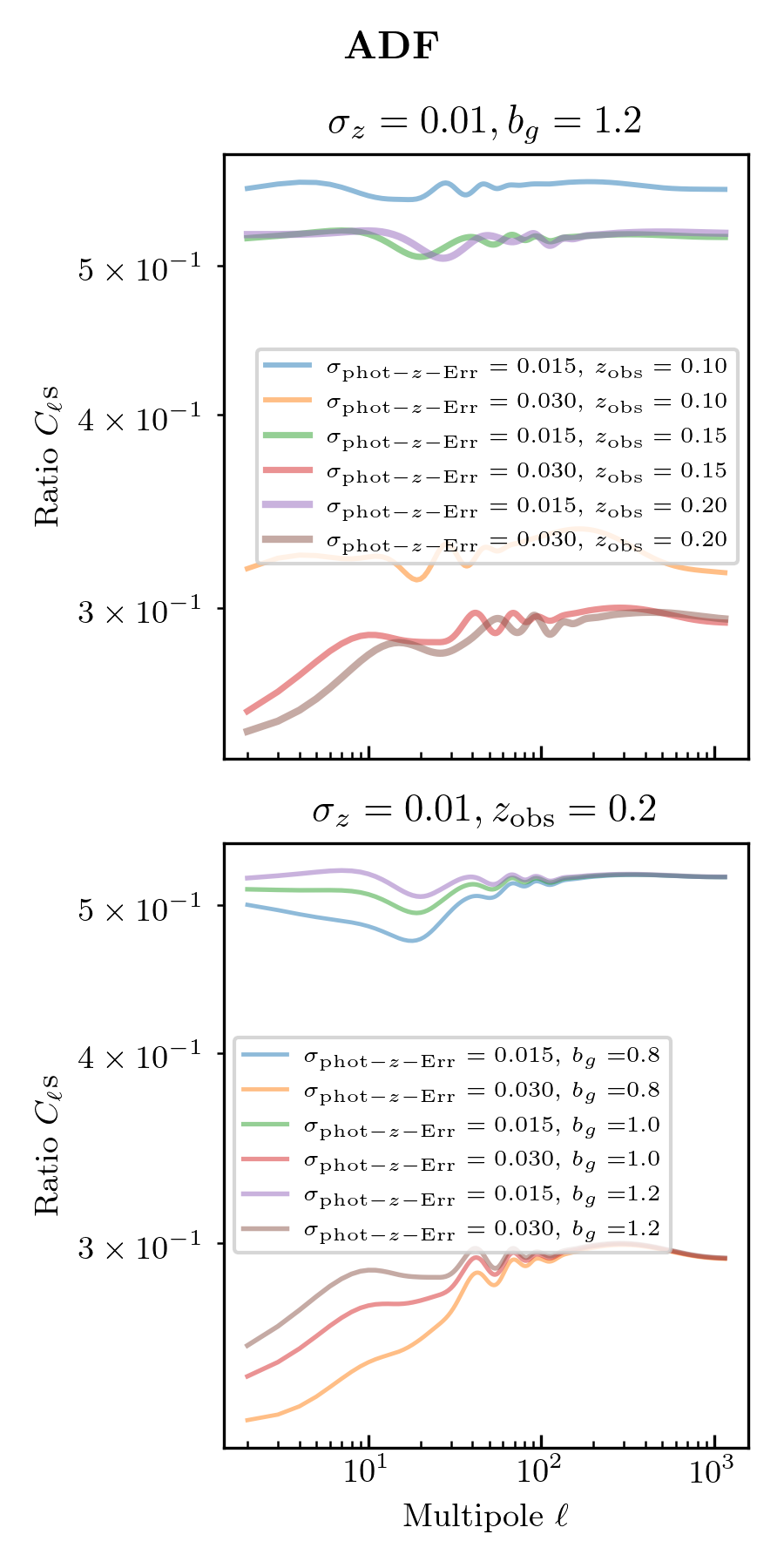}
   \includegraphics[width=0.4\paperwidth]{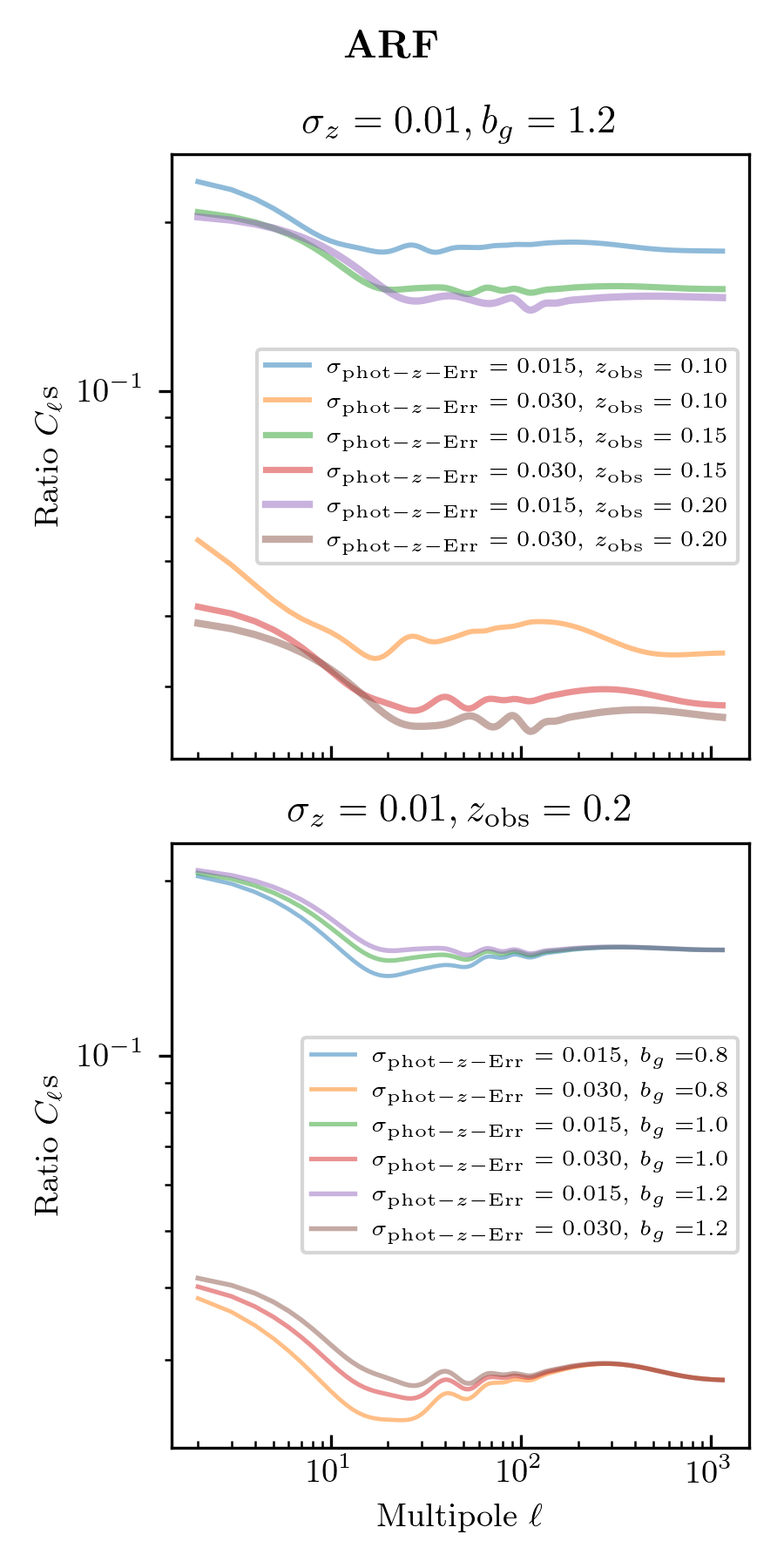}
      \caption{ {\it Left column:} Ratio of ADF $C_{\ell}$ with non-zero photo-$z$ errors wrt the corresponding case with $\sigma_{\rm Err}=0$. Different central redshift, shell width and linear bias configurations are displayed. {\it Right column:} Same as left column but for ARF. The suppression of the $C_\ell$s induced by the photo-$z$ errors is much more significant than for ADF. All $C_{\ell}$ computations have been conducted using {\tt ARFCAMB}. }
    \label{fig:ratioCells}
    \end{figure*}

%________________________________________________________________

Besides $b_g(z)$ and ${\cal A}_v$, there exists another parameter to be determined in our model that has not explicitly appeared yet. The effective width of the Gaussian redshift shells depends on the typical error in the photo-$z$ estimations of our galaxy samples. Boltzmann codes like {\tt ARFCAMB} require as part of the input the redshift distribution of the sources, i.e., $dN/dz$. In our tomographic study, this redshift distribution is then multiplied by a Gaussian window with parameters ($z_{\rm center}, \sigma_z$). But since the selection of our galaxies use the photo-$z$s, the actual width of our Gaussian window is the sum in quadrature of $\sigma_z$ with the typical photo-$z$ error, $\sigma_{\rm Err}$. The correct shape for $dN/dz$ and $(z-\bar{z})\, dN/dz$ to be input to our Boltzmann code hence must include a convolution with the typical photo-$z$ error PDF. Therefore the shape of this PDF and its width $\sigma_{\rm Err}$ are input parameters for our latest version of {\tt ARFCAMB}. 

In our analyses, and for the sake of simplicity, we have assumed that the photo-$z$ PDF shapes are Gaussian for all shells, with a width $\sigma_{\rm Err}$ to be determined. We may either assume that $\sigma_{\rm Err}$ is constant and the same for all redshifts, or search for a different value for every $z_{\rm center}$.

The inclusion of non-zero photo-$z$ errors results in a clear suppression of the amplitude of the $C_{\ell}$s for the ARF, while for the ADF this suppression is  more modest but not negligible, more clearly visible for shells of thinner or comparable width than $\sigma_{\rm Err}$.  Using {\tt ARFCAMB}, we have computed
the ratio of the ADF and ARF angular power spectrum multipoles of the non-zero $\sigma_{\rm Err}$ case over the no photo-$z$ error case, for different configurations of central redshift, shell width, bias, etc (see Fig.~\ref{fig:ratioCells}). In practice, we compute a grid of models for this ratio in terms of $z_{\rm center}$, $\sigma_z$, $b_g$, and $\sigma_{\rm Err}$, and interpolated through this grid when sampling $\sigma_{\rm Err}$.  

\subsection{The estimation of the covariance matrix}
\label{sec:covM}

Given the complexity in mimicking several properties of J-PLUS DR3 data (such as the {\it odds}-dependent selection function, or the impact of systematics residuals in auto- and cross-angular power spectra), we have opted to use the real data vector to build the covariance matrix rather than relying on a large number of approximated galaxy mocks. We further simplify this step by neglecting all non-Gaussian contributions to the covariance matrix. While our goal here is to constrain the galaxy bias, the photo-$z$ errors in our galaxy sample, and the amplitude of radial peculiar velocities in the local universe, we are not interested on to providing the tightest measurements possible with this data set. Rather, we focus our study on the compatibility of the ADF and ARF constraints, the possible impact by systematics of any type, and on whether it is possible or not to combine both probes to measure the same set of cosmological and astrophysical parameters.  

We thus adopt the standard Gaussian approximation for the covariance matrix. If we let latin indexes $i$ and $j$ denote values of pairs the central redshift and Gaussian widths $(z_{\rm center},\sigma_z)$, and provided we are considering in our data vector the {\rm auto} angular power spectra for each $i$, we can write the $(i,j)$ element of the covariance matrix as
\begin{equation}
\mathrm{Cov}^{\ell,\ell}_{i,j}= \langle C_{\ell}^{i,i} C_{\ell}^{j,j} \rangle - \langle C_{\ell}^{i,i} \rangle \langle C_{\ell}^{j,j} \rangle = \frac{2(C_{\ell}^{i,j})^2 }{(2\ell+1)f_{\rm sky}},
\label{eq:covM}    
\end{equation}
where the power spectra $C_{\ell}^{i,j}$ are obtained from real data. 
We are assuming the covariance matrix is block diagonal in the sense that we are neglecting that different multipoles ($\ell\neq\ell'$) are not independent. In fact, given that a binned multipole is written as $C_L = 1/N_L \sum_{\ell \in L} C_\ell$, it can be easily shown that the diagonal elements of the binned covariance matrix can be written as
\[
\mathrm{Cov}_{L,L} = \langle C_L C_L \rangle - \langle C_L \rangle \langle C_L \rangle 
\]
\[
= \frac{1}{N_L\times N_L} \sum_{\ell \in L} \sum_{\ell' \in L}\mathrm{Cov}_{\ell,\ell'} 
\]
\begin{equation}
\phantom{xxxxxxxxxx} 
= \frac{1}{N_L} \sum_{\ell \in L}  \mathrm{Cov}_{\ell,\ell} + \mathrm{ND}\frac{N_L-1}{N_L}. 
\label{eq:covM_id}
\end{equation}
In this equation, $\mathrm{ND}$ is the average amplitude of all off-diagonal elements of the covariance matrix inside the $L$ multipole bin. The first term in the right hand side of the equation is simply the average of the diagonal elements of $\mathrm{Cov}_{\ell,\ell'}$ along the $\Delta \ell=10$ bin. We are thus neglecting the last term, and this should yield to lower covariances and correspondingly lower errors in our inferred parameters. While we should keep this in mind when interpreting our results, we should also note that this approach should contain all (Gaussian) contributions to the diagonal of the covariance matrix from systematics, foregrounds, and non-linearities present in the data, which are not straightforward to model.

\begin{figure}
   \centering
   \includegraphics[width=0.4\paperwidth]{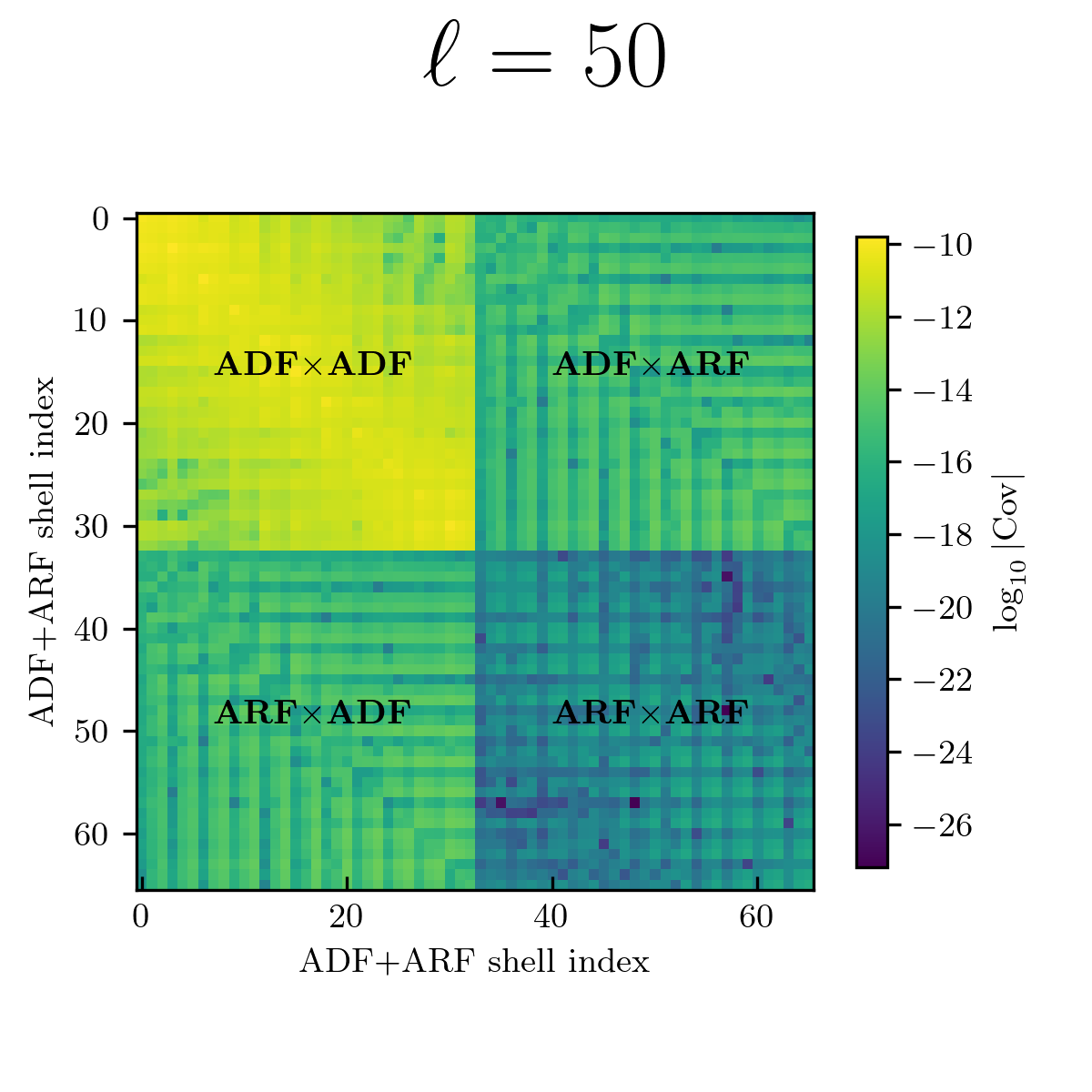}
      \caption{Logarithmic 2D map of the absolute value of the covariance matrix (for $\ell=50$) when considering both the ADF and the ARF probes. The first block corresponds to the ADF, while the second to the ARF (with typically lower amplitudes than the ADF). The matrix index first runs over $\sigma_z$, and then over $z_{\rm centre}$. It can be see that the narrowest shells for ARF have lower amplitude entries.
              }
    \label{fig:covM}
    \end{figure}

In Fig.~\ref{fig:covM} we show a 2D plot of the correlation  matrix when considering both ADF and ARF probes.

%-------------------------------------------------------------
%-------------------------------------------------------------

\subsection{Constraining the parameters}
\label{sec:par_const}
We implement a Monte Carlo Markov Chain (MCMC) algorithm \citep[using the software {\tt emcee},][]{emcee}) to constrain two different parameter sets under to different configurations:
\begin{itemize}
    \item First parameter configuration given by $\{\sigma_{\rm Err}, b_g^i, {\cal A}_v\}$, where the superscript $i$ runs through all 11 redshift shells ranging from $z_{\rm center}=0.05$ up to $z_{\rm center}=0.25$, and where $\sigma_{\rm Err}$ is taken to be the same for all redshift shells.
    \item A second parameter configuration extending the previous one to $\{ \sigma_{\rm Err}^i,  b_g^i, {\cal A}_v \}$, where we let the typical uncertainty in the photo-$z$ estimates to be different for each redshift shell. 
\end{itemize}

The uniform priors assigned to the parameters are $[0,0.026]$ (first parameter configuration) and $[0,0.0325]$ (second parameter configuration) for $\sigma_{\rm Err}^i$ (the maximum values is taken to be twice an 2.5 times the value provided by {\tt LePhare}, respectively), and $[0,5]$ for both the $b_g^i$s and ${\cal A}_v$. These ranges are sufficiently wide so that if some of the parameters under search hits the prior upper limits one should invoke either non-linear physics, systematics, or a combination of both. We adopted 8 different walkers per parameter under consideration (so 104 walkers for the first parameter configuration, and 184 for the second one), with 3,000 steps per walker, with a burn-out phase of 400 (which was checked to be long enough for convergence).

\section{Results}
In this section we describe the results of fitting the ADF and ARF angular power spectra with the linear model described in the previous section. In order to test for consistency, we first conduct and interpret the ADF and the ARF tomography independently of each other, and compare the constraints obtained in either case. We next opt to combine both probes, and extract joint parameter constraints from those, even if in some cases the ADF and ARF outputs are incompatible and their combination must be interpreted carefully.  
\label{sec:results}

\begin{figure*}
\centering
\hspace*{-1.cm}
\includegraphics[width=0.27\paperwidth]{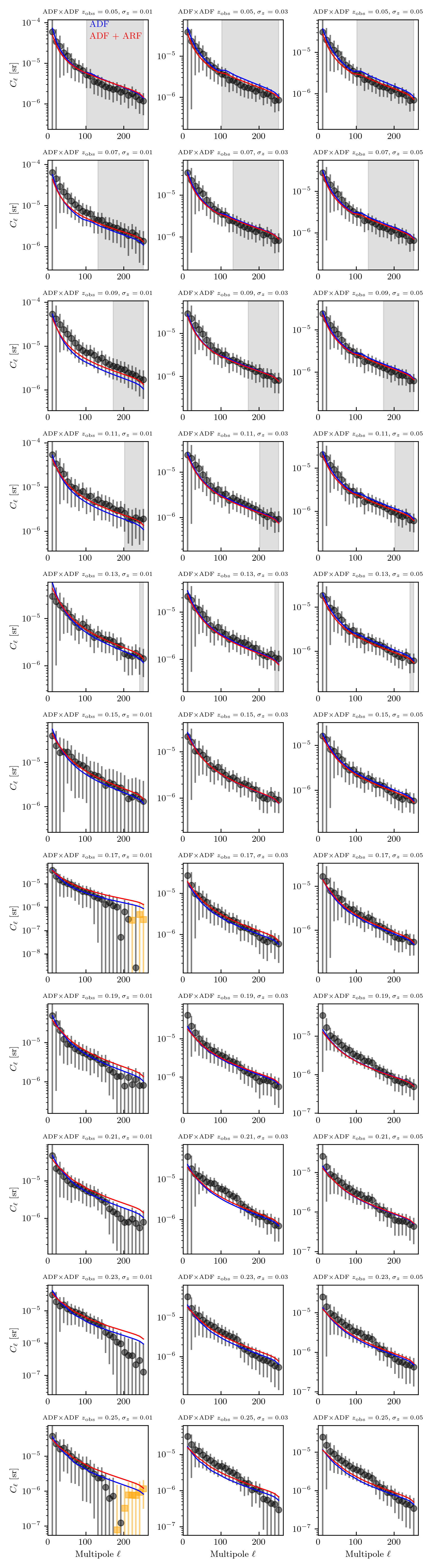}
\hspace*{1.cm}
\includegraphics[width=0.27\paperwidth]{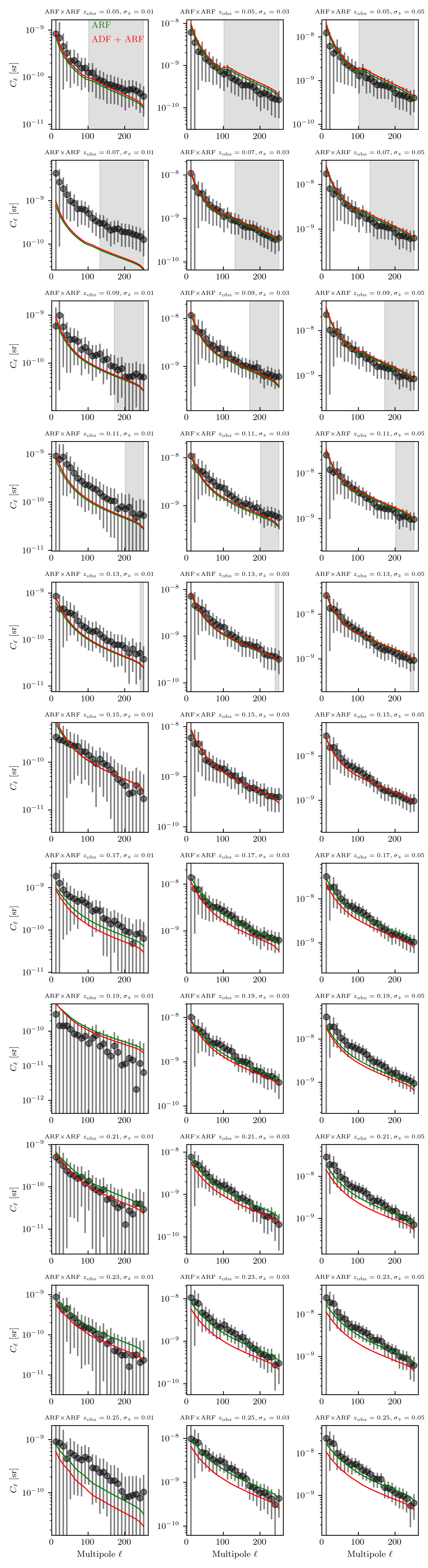}
  \caption{ {\it Left column:} Observed (binned) angular power spectra for each central redshift (running rows) and Gaussian width (running in columns). These $C_{L}$s are {\em not} corrected by the mask, but are corrected by our estimation of shot noise. In some few cases this shot noise prediction is larger than the observed power spectrum power bands, yielding negative values displayed by orange squares. The gray regions visible in top rows display the angular scales projecting below the non-linear threshold of $5~h^{-1}$~Mpc at each redshift. The solid lines display the best-fit linear models, in blue when considering ADF only, in red when considering ADF$+$ARF.  {\it Right column:} Same as left column but for ARF. %The green solid lines display the best-fit solution when considering ARF only. 
  }
\label{fig:cls_obs}
\end{figure*}

In Fig.~\ref{fig:cls_obs} we display, for the BANNJOS catalogue, the observed angular power spectra for each redshift shell (rows) and each shell width (columns) for ADF (left panel) and ARF (right panel). The range of multipoles in those panels that correspond to projected distances below 5~$h^{-1}$~Mpc are depicted in gray, and excluded from the fits. These angular power spectra have been corrected for shot noise, but not for the mask. The solid curves provide the best-fit theoretical model after convolving by the effective sky mask. In the left panel referring to ADF, the blue curves denote the best fit obtained after including only ADF in the MCMC chains (and assuming a constant value of the photometric redshif error rms $\sigma_{\rm Err}$ for all redshift shells). The red curve, instead, considers both ADF and ARF data in the parameter estimation. Analogously, in the right panel displaying ARF angular power spectra, the green curve provide best-fit models for ARF data only, and the red ones for ADF$+$ARF. 
%________________________________________________________________

%________________________________________________________________

Overall the linear model adopted here provides a fairly good fit to the data in most occasions. Narrow shells (displayed in the left columns of both panels) contain fewer galaxies, and suffer more from shot-noise residuals. These shells for ARF are noisier than for ADF, since they contain line-of-sight gradients that carry more uncertainty. Narrow shells should also be impacted more by non-linearities (as it was found in \citet{arf_letter1}), provided they are comparatively sensitive to smaller scales than wider shells. In some cases there is over-correction for shot noise (negative values of the $C_{\ell}$s are given by orange squares), but this is restricted to high multipoles. Since fits for the bias parameter are done separately for each redshift shell (but for the three widths at the same time), wider shells containing more galaxies typically carry most of the weight: if there exists tension between the data and the linear model, then narrow shells are the ones most likely showing some degree of discrepancy. This tension for narrow shells is more evident for ARF than for ADF, although for ARF error bars are significantly larger than for ADF, leaving room for optimal fits far from the narrow shell data.  

\begin{figure*}
\centering
%\hspace*{-1.cm}
\includegraphics[width=0.8\paperwidth]{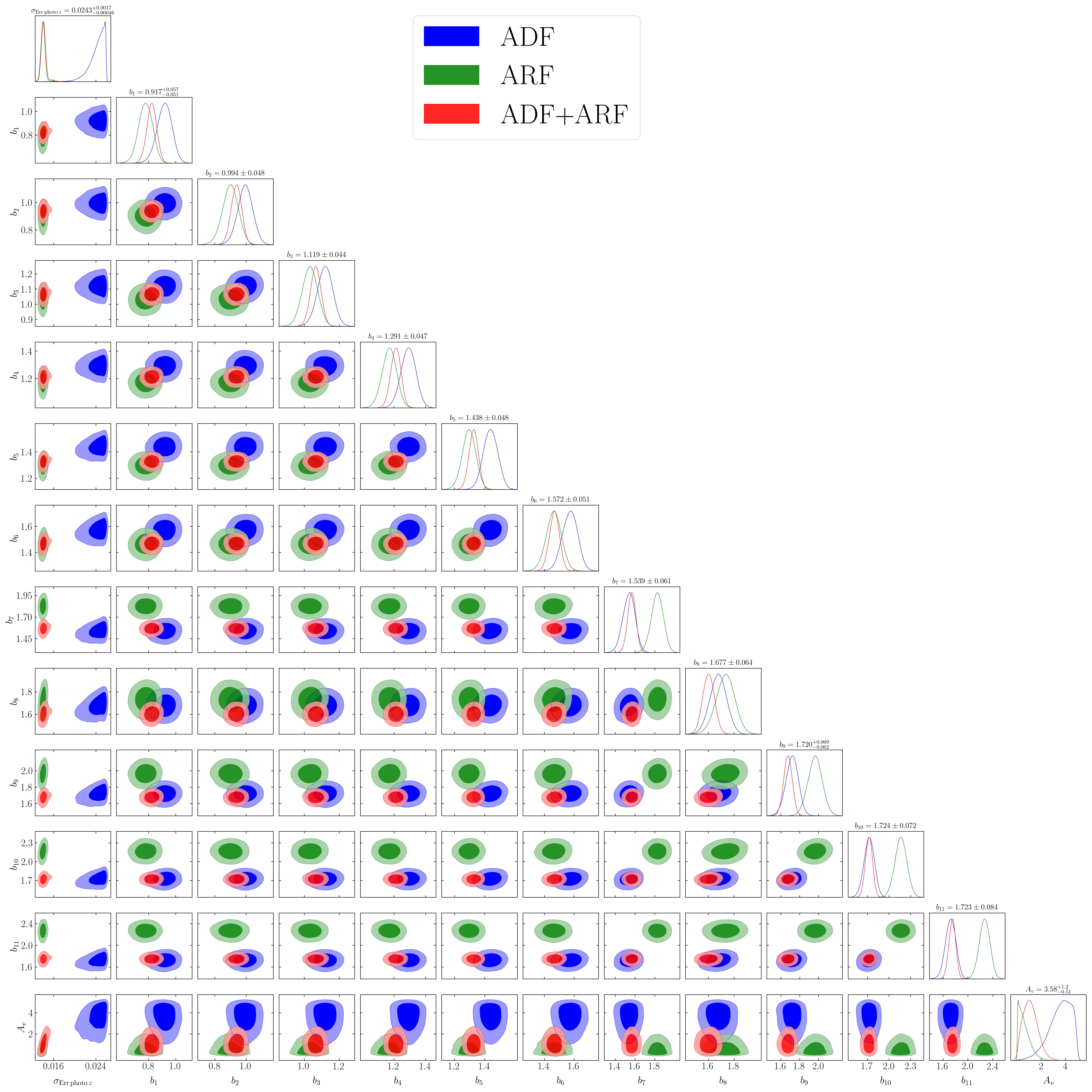}
\caption{Corner plot for the first parameter configuration introduced in Sect.~\ref{sec:par_const} where $\sigma_{\rm Err}$ is assumed constant for all redshifts. The parameter set is $\{\sigma_{\rm Err}, b_g(z_i), {\cal A}_v\}$, with the subscript $i$ running from $1$ to the number of redshift shells $N_{\rm shells}=11$. The galaxy catalogue under use is BANNJOS. }
\label{fig:mcmc_2c}
\end{figure*}

\begin{figure*}
\centering
%\hspace*{-1.cm}
\includegraphics[width=0.8\paperwidth]{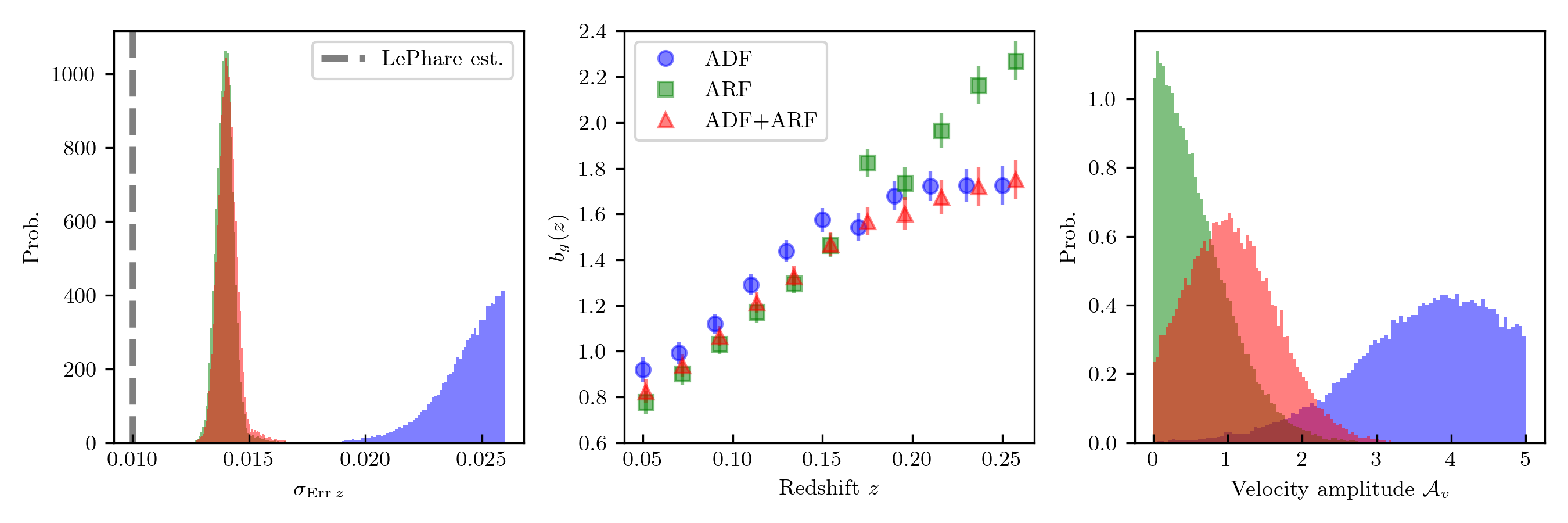}
\caption{Projected results for the first parameter configuration in Sect.~\ref{sec:par_const} where $\sigma_{\rm Err}$ is assumed constant for all redshifts. {\it (Left panel):} Posterion probability density distribution for $\sigma_{\rm Err}$, the average uncertainty in the photo-$z$ error. The value provided by {\tt LePhare} is $\sigma_{\rm Err}^{\rm LePhare}=0.010$, significantly lower than the ARF and ADF$+$ARF constraints. {\it (Middle panel):} Constraints on the linear bias paramter $b_g(z)$ for each redshift shell. {\it (Right panel):} Constraints in the relative radial peculiar velocity amplitude with respect to the fiducial model ${\cal A}_v$. The galaxy catalogue under use is BANNJOS.}
\label{fig:results_2c}
\end{figure*}

The joint ADF$+$ARF fits (red curves) are in many cases indistinguishable from the ADF- and ARF-only fits, particularly at low redshifts. Some tension between the 1- and 2-probe fits starts to appear at $z\gtrsim 0.17$, at a higher level for ARF than for ADF. Nevertheless, similar best-fit curves for one- and two-probes do not necessarily involve similar best-fit parameters. Indeed, as can be seen in Fig.~\ref{fig:mcmc_2c} for the case where $\sigma_{\rm Err}$ is constant for all redshifts and the BANNJOS selection, the ADF-only and ARF-only contours (blue and green colours, respectively) are clearly incompatible for $\sigma_{\rm Err}$ (the first column of this plot corresponds to this parameter). The tension between the bias contours grows with redshift, surpassing the $\sim 2-\sigma$ level by first time at $z=0.17$, and then again at $z=0.23$ and $z=0.25$. The bias values from ADF at $z<0.17$ are systematically higher than those from ARF, but at the same time, $\sigma_{\rm Err}$ is also much higher. These two parameters are obviously correlated, since higher values of the bias $b_g(z)$ combined with higher values of $\sigma_{\rm Err}$ can yield similar amplitudes of the $C_{\ell}$s than those obtained under lower values of these two parameters, provided that each of them have opposite effect on the $C_{\ell}$ amplitudes. Again for the BANNJOS catalogue, Fig~\ref{fig:results_2c} shows the projected best-fit values for the parameters under the first parameter configuration (constant $\sigma_{\rm Err}$). The joint ADF$+$ARF fit (in green colour) is driven by the ARF on $\sigma_{\rm Err}$ (since ARF are significantly more sensitive to this parameter), and also for bias parameters below $z<0.19$. Above this redshift the ADF$+$ARF values are dominated by the ADF. The value provided for $\sigma_{\rm Err}$ by the photo-$z$ code {\tt LePhare} is $\sigma_{\rm Err}=0.010$, which is below (at the $\simeq 9~\sigma$ level) the distribution for this parameter from ARF, and also clearly discrepant with the value provided by ADF. Regarding the velocity amplitude ${\cal A}_v$, there is a significant mismatch between the ADF and the ARF outputs: ARF see no evidence for radial peculiar velocities (${\cal A}_v=0.42\pm 0.57$), ADF point to ${\cal A}_v=3.4\pm 1.0$, while the joint ADF$+$ARF fit provides an intermediate solution that just by coincidence peaks around ${\cal A}_v\sim 1$, ${\cal A}_v=0.90\pm 0.56$. 
%________________________________________________________________
%________________________________________________________________
%________________________________________________________________

\begin{figure*}
\centering
%\hspace*{-1.cm}
\includegraphics[width=0.8\paperwidth]{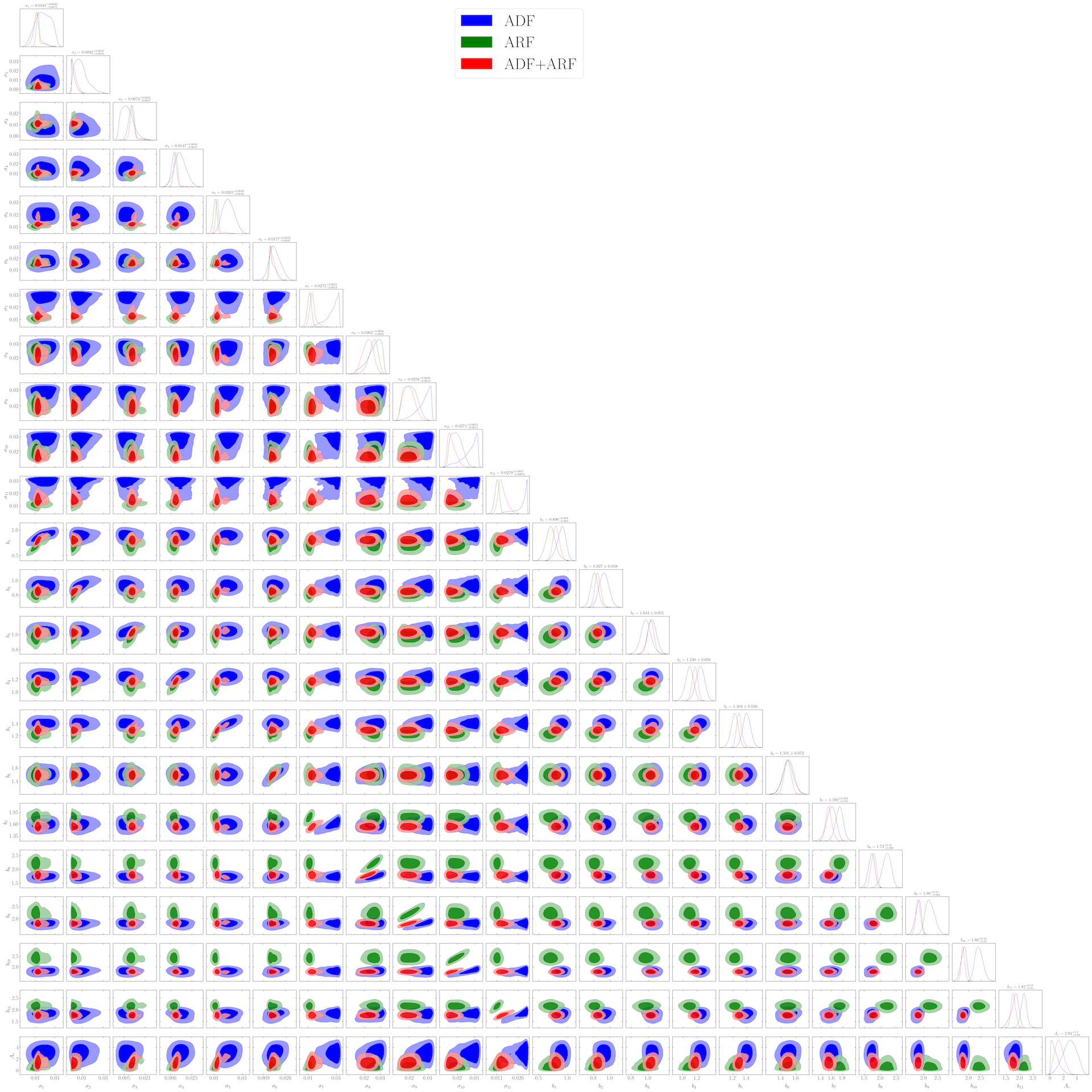}
\caption{Corner plot for the second parameter configuration introduced in Sec.~\ref{sec:par_const}, consisting in the parameter set $\{\sigma_{\rm Err}^i,b_g(z_i),{\cal A}_v\}$, where the subscript $i$ runs from $1$ to the number of redshift shells $N_{\rm shells}=11$. The galaxy catalogue under use is BANNJOS.}
\label{fig:mcmc_6c}
\end{figure*}

\begin{figure*}
\centering
%\hspace*{-1.cm}
\includegraphics[width=0.8\paperwidth]{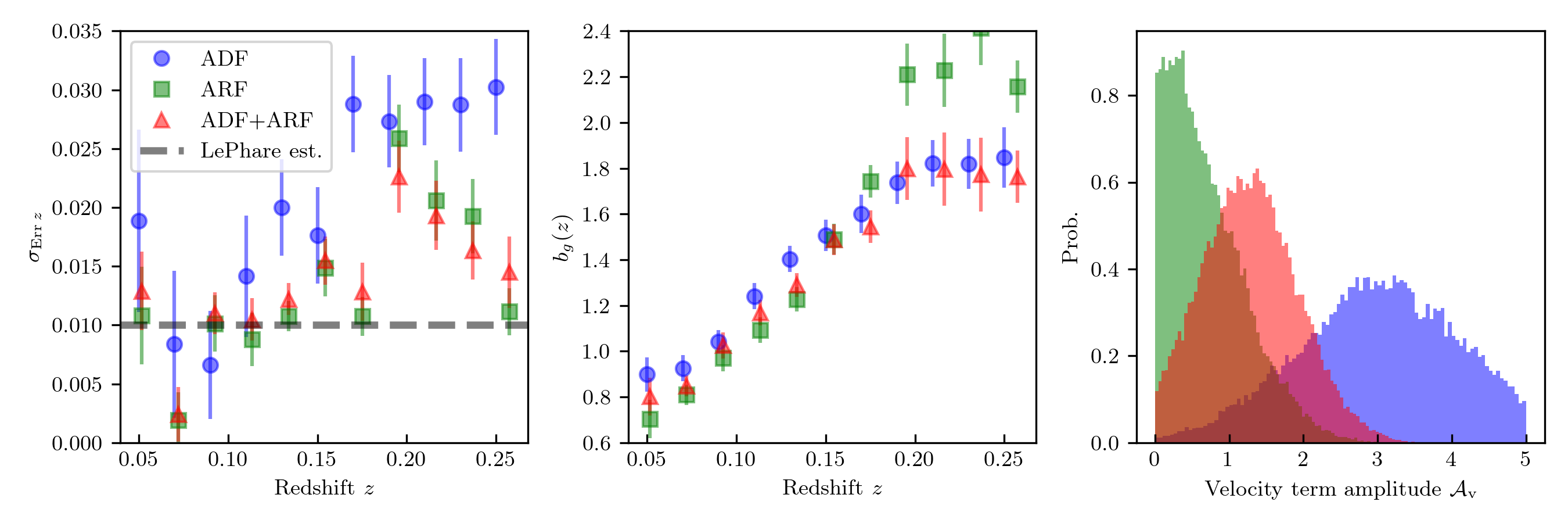}
\caption{Same as Fig.~\ref{fig:results_2c} but for the second parameter configuration of Sec.~\ref{sec:par_const} observing a parameter set given by $\{\sigma_{\rm Err}^i,b_g(z_i),{\cal A}_v\}$, with $i$ running from $1$ to the number of redshift shells $N_{\rm shells}=11$. The galaxy catalogue under use is BANNJOS.}
\label{fig:results_6c}
\end{figure*}

If we instead look at the results from the second configuration ($\sigma_{\rm Err}$ as a function of redshift), we find again for the BANNJOS catalogue that there is actually agreement for $\sigma_{\rm Err}$ at low redshifts between both probes, although systematically higher for ADF than for ARF, which again translates into higher ADF bias estimates. This is shown in the corner plot Fig.~\ref{fig:mcmc_6c}, whose results are summarised in Fig.~\ref{fig:results_6c}. A careful look at Fig.~\ref{fig:mcmc_6c} reveals a positive correlation of the $\{ \sigma_{\rm Err}(z), b_g(z) \}$ parameters for the same shells: this is found in a diagonal whose first panel is located in the first column and 12th row, displaying the correlation between $\sigma_{\rm Err}(z)$ and $b(z)$ for the first redshift bin. This diagonal follows further in the 13th row and second column. In Fig.~\ref{fig:results_6c} one can also notice that it is only at the highest redshift shells ($z>0.15$) when $\sigma_{\rm Err}$ estimates for ADF reach the $\sim 0.025$ level, which is higher than any ARF $\sigma_{\rm Err}$ estimate. Thus, on what regards $\sigma_{\rm Err}$ and $b_g(z)$, ADF and ARF agree for $z<0.17$, but it must be stressed that ARF provide much tighter constraints on these two parameters, very significantly shrinking their degeneracy area.

%________________________________________________________________
%________________________________________________________________
%________________________________________________________________

\section{Discussion}
\label{sec:discussion}

Despite of the modest cosmological volume probed by J-PLUS DR3, and the correspondingly small scales sampled by the survey, the linear model seems to provide a surprinsingly good fit to the data. The agreement, for most shells and widths, of the ADF-/ARF-only fits and the joint ADF$+$ARF fits is also remarkable. A visual analysis in Fig.~\ref{fig:cls_obs} of the observed angular power spectra suggests that non-linearities impact more ARF than ADF, particularly on the narrower shells ($\sigma_z=0.01$). This was found in \citet{arf_letter1} while comparing the linear theory predictions for the ARF angular power spectra with the average outcome of 100 COLA\footnote{The {\it Comoving Lagrangian Acceleration} algorithm \citep{COLA} permits producing N-body catalogues of dark matter particles under any input cosmology with high precision up to relatively small scales with a fraction of the CPU cost of the corresponding full N-body simulation.} simulations: while the fits in real space were acceptable, the simulations provided a systematically $\sim 5~\%$ lower amplitude for all terms containing radial velocity contributions. This mismatch was modelled by adding a thermal, random velocity field component in the model. However, in this case, the measured ARF for narrow shells is in most cases above the linear-theory best fit. This power excess seems to be more attributable to the non-linear density contrast, which, however, does not show up to the same extent in the corresponding cases for ADF (although we do see a systematic higher amplitude of the measured $C_{\ell}$s above the best-fit model for most of these narrow shells).   

Even when the best-fit angular power spectra for ADF-/ARF-only and ADF$+$ARF
are very similar, the inferred values for $b_g(z)$ and $\sigma_{\rm Err}$ are not. This is caused by the aforementioned degeneracy between these two parameters, since they impact the amplitude of the $C_{\ell}$s in opposite ways. The recovered values for $\sigma_{\rm Err}$ and ADF are (in all redshift bins but one) above the corresponding values for ARF (but still compatible with these) if $z<0.17$. This translates in correspondingly higher ADF values for $b_g(z)$ (but again compatible to those from ARF if $z<0.17$, see the corresponding PDF panels in Fig.~\ref{fig:mcmc_6c}). At higher redshifts ($z>0.17$) the ADF $\sigma_{\rm Err}(z)$ values climb up to $\sigma_{\rm Err}(z)\sim 0.025-0.03$, which is above twice the average value provided by {\tt LePhare}. Since these high-$z$ redshift shells carry significant statistical weight, they drive the $\sigma_{\rm Err}$ estimate to $\sim 0.025$ when it is considered as constant throughout redshift shells (first parameter configuration, left panel in Fig.~\ref{fig:results_2c}). 

%________________________________________________________________
\begin{figure*}
\centering
%\hspace*{-1.cm}
\includegraphics[width=0.8\paperwidth]{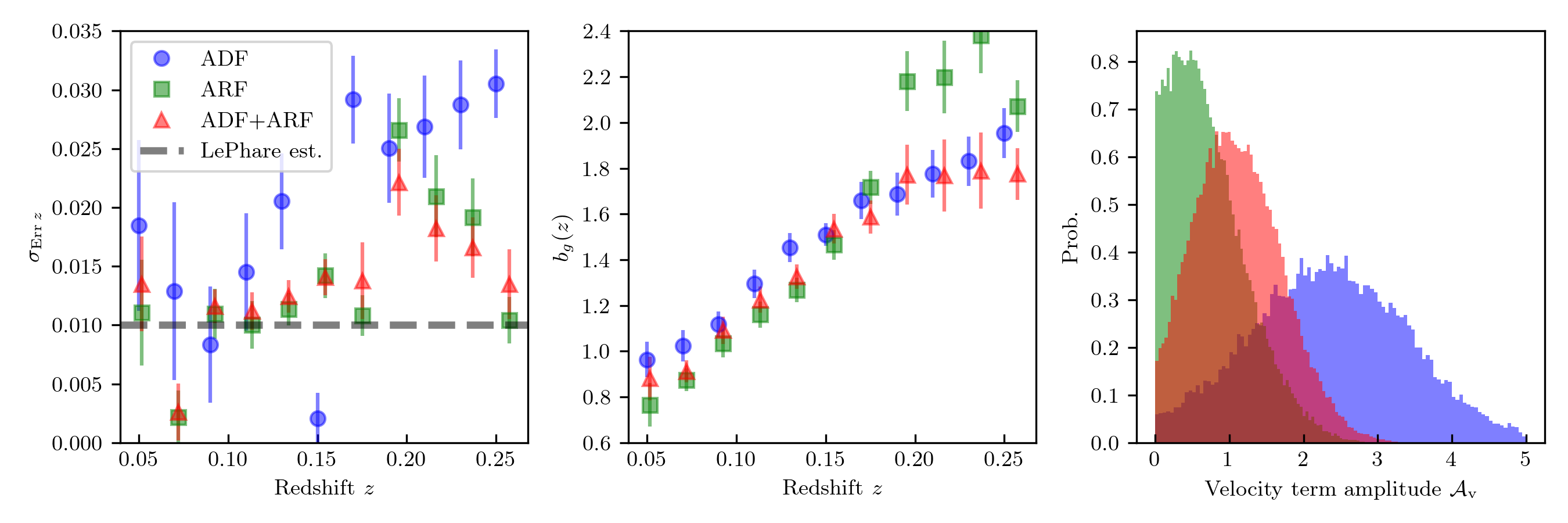}
\caption{Same as Fig.~\ref{fig:results_6c} but for the XGBoost galaxy catalogue}
\label{fig:results_6c_RvM}
\end{figure*}
%________________________________________________________________

Surprinsingly, for $z>0.17$, the ADF bias estimates flatten in a plateau at $b_g(z)\sim 1.7$ while the ARF counterparts keep increasing monotonically with redshift up to $b_g(z)\sim 2.3$ for $z>0.23$. One would naively expect the bias to increase with redshift, or at least this is the behavior expected for halos at fixed mass. In our case the selection is done under a cut in both apparent $r$-band magnitude and $odds$, and this suggests that higher redshift objects should correspond to more luminous (and thus more massive) haloes, which are bound to expectations for relatively higher bias values. Since the best-fit models for ADF still fit the data remarkably well for those high redshift shells, the low bias estimates must be necessarily balanced by the contribution of the velocity term. The velocity amplitude estimates for ADF are well above the expected value of unity, ${\cal A}_v\sim 2-3$, whereas ARF estimates peak around ${\cal A}\sim 0$ and exclude values above 3 with very high significance. The velocity term contributes preferentially at low and intermediate multipoles, which is also the angular regime where systematics tend to contribute the most (see Fig.~\ref{fig:cls_corr}). A likely interpretation of the parameter constraints is thus that systematics residuals significantly contribute to these high-redshift shells, triggering unphysical high values of ${\cal A}_v$ which leaves room for moderately low values of $b_g(z)$ while keeping $\sigma_{\rm Err}(z)$ high. Since these high redshift bins have a significant statistical weight (the associated errors in the $C_{\ell}$s are relatively small provided no scale cut is applied and shot noise is similar than in lower redshift bins), they tend to drive the joint ${\cal A}_v$ estimation to high values ( ${\cal A}_v>1$). 

%________________________________________________________________
\begin{figure}
\centering
%\hspace*{-1.cm}
\includegraphics[width=0.4\paperwidth]{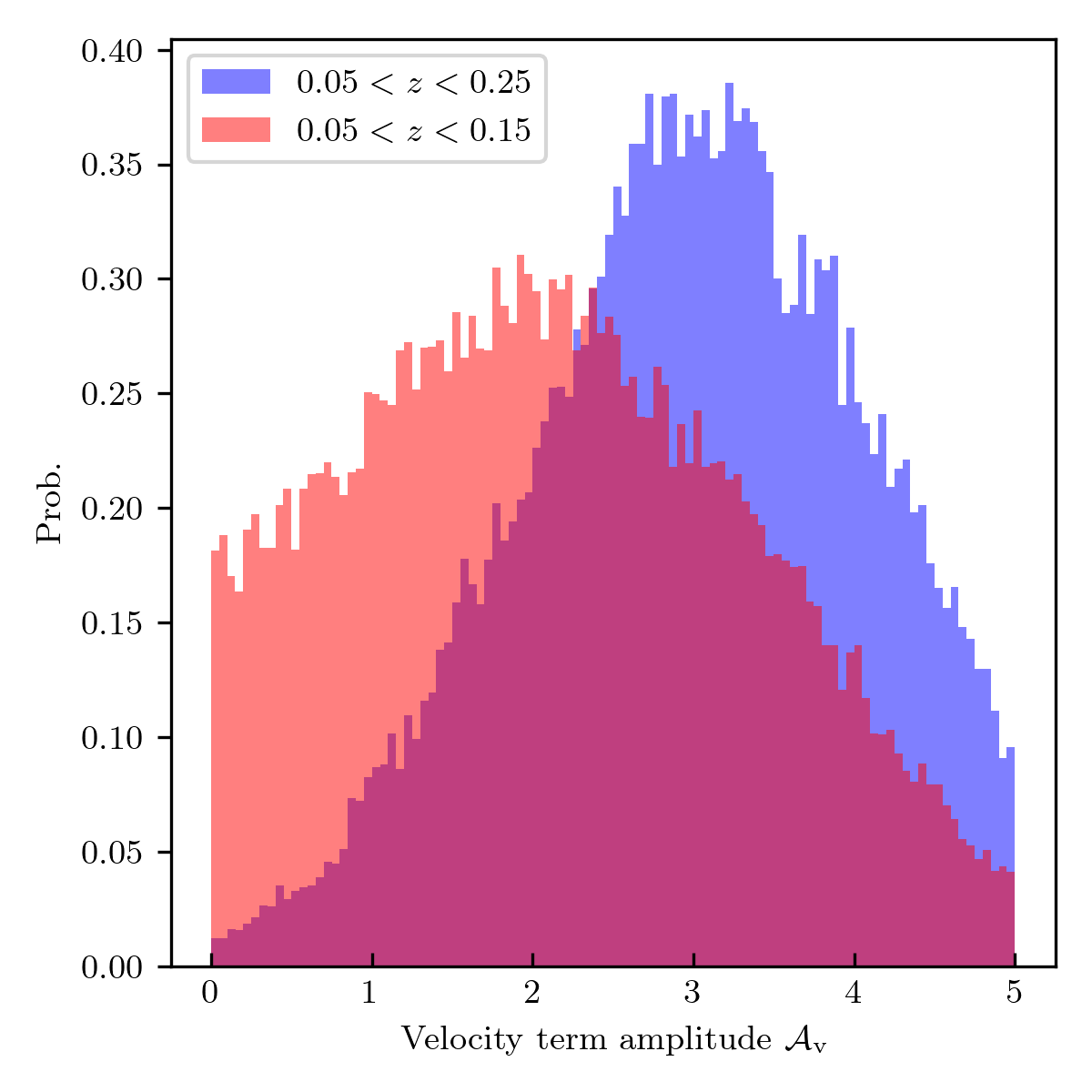}
\caption{Distribution of the recovered ${\cal A}_v$ values for $0.05<z<0.15$ (red) and $0.05<z<0.25$ (blue), for the second parameter configuration (with redshift-dependent $\sigma_{\rm Err}$). The galaxy catalogue under use is BANNJOS.}
\label{fig:A_vs}
\end{figure}
%________________________________________________________________

We further test the stability of these results by comparing them with the outcome of the XGBoost catalogue given in Fig.~\ref{fig:results_6c_RvM}, to be compared to Fig.~\ref{fig:results_6c}. Practically all estimates of $\sigma_{\rm Err}(z)$ and $b_g(z)$ remain compatible (given their uncertainties) in both catalogues, with the only exception for $\sigma_{\rm Err}(z)$ at $z=0.15$, which for ADF becomes $\sigma_{\rm Err}(z=0.15)\simeq 0.002 \pm 0.002$, in tension with its value for the BANNJOS catalogue ($\sigma_{\rm Err}(z=0.15)\simeq 0.017 \pm 0.005$). This pronounced variation at the $3-\sigma$ level partially reflects the optimistic approach for the covariance matrix adopted in Sect.~\ref{sec:covM}. We nevertheless conclude that overall this comparison points towards a compatible parameter estimation obtained for either galaxy catalogue.

\section{Conclusions}
\label{sec:conclusions}

In this work we have conducted a tomographic study of the angular density and redshift fluctuations in a high $odds$ ($odds>0.8$), relatively deep ($r<21$) galaxy sample from J-PLUS DR3. Our analyses have shown that both ADF and ARF are able to yield measurements of the galaxy bias $b_g(z)$ and the photo-$z$ error $\sigma_{\rm Err}(z)$ for each redshift shell. ADF and ARF provide also constraints on the amplitude of radial peculiar velocities ${\cal A}_v$, although these are incompatible for reasons we address below. Overall, a simplistic linear-theory model with two sets of redshift bin parameters ($\{\sigma_{\rm Err}(z),b_g(z)\}$) and an overall radial peculiar velocity amplitude ${\cal A}_v$ suffice to provide a surprinsingly good fit to the data. The results we obtain do not depend significantly on which of the two galaxy catalogues (BANNJOS or XGBoost) is under analysis: both yield compatible results. 

The estimates of $\sigma_{\rm Err}(z)$ are compatible from either the ADF or the ARF probe for $z<0.17$ (but almost always higher for ADF than for ARF), a redshift range where we show systematics to have a low impact on the angular distribution of galaxies. These estimates of $\sigma_{\rm Err}(z)$ scatter around $\sigma_{\rm Err}(z)=0.014$ with a typical uncertainty of about 30~\%, while the estimate from {\tt LePhare} is $\langle \sigma_{\rm Err}(z)\rangle=0.010$. ARF are significantly shrinking the degeneracy area between $\sigma_{\rm Err}$ and $b_g(z)$, and this translates into more precise estimates of these parameters. On that same redshift range, the galaxy bias estimates from both ADF and ARF seem to grow monotonically with $z$, with a very similar slope, from $b(z)\simeq 0.7$ at $z=0.05$ up to $b(z)\simeq 1.6$ at $z=0.15$. Although compatible, ADF bias estimates again lie systematically above ARF ones, and this is understood given the correlation between $b_g(z)$ and $\sigma_{\rm Err}(z)$ for any given redshift shell: higher values of these two parameters impact the amplitude of the angular power spectrum in opposite ways, and it is thus possible to leave the amplitude of the $C_{\ell}$s practically untouched after increasing both parameters conveniently. 

While ARF see no evidence for non-zero ${\cal A}_v$, the ADF measurements of this parameter are clearly discrepant, since they point to ${\cal A}_v \approx 3.0\pm 1.0$. ADF are known to be less sensitive than ARF to velocity terms and thus to ${\cal A}_v$ \citep[see, e.g.,][]{arf_letter1,Legrand_ARF}, and for $0.05<z<0.15$ ADF constraints on ${\cal A}_v$ are almost flat in the entire range ${\cal A}_v \in [0,5]$. In this redshift range, ARF exclude high values of ${\cal A}_v$ ( ${\cal A}_v>3$) at very high significance. When higher redshift bins are included in the analysis, the ADF bias estimates $b_g(z)$ flatten out (contrary to the corresponding ARF estimates, which keep increasing monotonically with redshift), and $\sigma_{\rm Err}(z)$ from ADF also take their highest values. At the same time, ADF-derived constraints of ${\cal A}_v$ tend to prefer high (${\cal A}_v\approx 3$) values. We interpret these high ${\cal A}_v$ values as the reaction of the linear model to accommodate low $b_g(z)$ and low $\sigma_{\rm Err}(z)$ values and some likely systematics residual in the low-to-intermediate $\ell$ domain where the velocity-related terms have larger impact. 

But for the lower redshift bins ($z<0.15$), ours are (to our knowledge) the first constraints on the photo-$z$ errors $\sigma_{\rm Err}$ derived from galaxy ADF and ARF tomography: by measuring angular power spectra and comparing them to theoretical expectations, we are able to derive estimates of $\sigma_{\rm Err}$ that are similar to (although higher by about $\sim 40~\%$) the value provided by the photo-$z$ estimator ({\tt LePhare}, $\sigma_{\rm Err}^{\rm LePhare}=0.010$). 

In future spectro-photometric surveys like J-PAS we expect to handle remarkably smaller photo-$z$ errors that should enable measuring the $b_g(z)$ and ${\cal A}_v$ parameters with little sensitivity to the uncertainties on $\sigma_{\rm Err}$. The first few hundreds of square degrees of area being currently covered by J-PAS constitute our next test bench where confronting those expectations with real data.

%________________________________________________________________

\section*{acknowledgements}
C.H.-M. acknowledges the hospitality of the Centro de Estudios de Física del Cosmos de Aragón (ceFca), where part of this work took place.
 C.H.-M. also acknowledges the support of the Spanish Ministry of Science and Innovation projects PID2021-126616NB-I00, PID2022-142142NB-I00, the European Union through the grant ``UNDARK'' of the Widening participation and spreading excellence programme (project number 101159929), and the contribution from the IAC High-Performance Computing support team and hardware facilities. VM thanks CNPq (Brazil) and FAPES (Brazil) for partial financial support. ET acknowledges funding from the HTM (grant TK202), ETAg (grant PRG1006) and the EU Horizon Europe (EXCOSM, grant No. 101159513). AE acknowledges the financial support from the Spanish Ministry of Science and Innovation and the European Union - NextGenerationEU through the Recovery and Resilience Facility project ICTS-MRR-2021-03-CEFCA. Based on observations made with the JAST80 telescope at the Observatorio Astrofísico de Javalambre (OAJ), in Teruel, owned, managed, and operated by the Centro de Estudios de Física del Cosmos de Aragón. We acknowledge the OAJ Data Processing and Archiving Unit Department\citep[DPAD,][]{davidCH12} for reducing and calibrating the OAJ data used in this work.
Funding for the J-PLUS Project has been provided by the Governments of Spain and Aragón through the {\it Fondo de Inversiones de Teruel}; the Aragón Government through the Research Groups E96, E103, E16\_17R, E16\_20R, and E16\_23R; the Spanish Ministry of Science, Innovation and Universities (MCIN/AEI/10.13039/501100011033 y FEDER, Una manera de hacer Europa) with grants PID2021-124918NB-C41, PID2021-124918NB-C42, PID2021-124918NA-C43, PID2021-124918NB-C44, PGC2018-097585-B-C21 and PGC2018-097585-B-C22; the Spanish Ministry of Economy and Competitiveness (MINECO) under AYA2015-66211-C2-1-P, AYA2015-66211-C2-2, AYA2012-30789, and ICTS-2009-14; and European FEDER funding (FCDD10-4E-867, FCDD13-4E-2685). The Brazilian agencies
FINEP, FAPESP, and the National Observatory of Brazil have also contributed to this project. Some of the results in this paper have been derived using the HEALPix \citep{healpix} package. 

%%%%%%%%%%%%%%%%%%%%%%%%%%%%%%%%%%%%%%%%%%%%%%%%%%
\section*{Data Availability}

 All original J-PLUS data can be accessed at \url{https://archive.cefca.es/catalogues}. The software implementing the analysis presented in this work can be accessed via direct request to the first author, and will be eventually accessible in the github site \url{https://github.com/chmATiac}.



% Bibtex style:------------
\bibliographystyle{mnras}
%\bibliography{references_pw,pip13II,refs_kDolagI,refs_kDolagII,Planck_bib}
\bibliography{biblio}
%--------------------------

\end{document}